\documentclass[useAMS,usenatbib]{mn2e}
\usepackage{graphics}
\usepackage{epsfig}
\title[Orbital Structure of Merger Remnants]{Orbital Structure of Collisionless Merger Remnants: On 
the Origin of Photometric and Kinematic Properties of
Elliptical and S0 Galaxies} 
\author[R. Jesseit, T. Naab and A. Burkert]{R. Jesseit$^{1}$\thanks{E-mail:jesseit@usm.uni-muenchen.de}, 
T.Naab$^{2}$\thanks{E-mail: naab@usm.uni-muenchen.de} \thanks{Present Address: Universit\"atssternwarte, Scheinerstr. 1, 
81679 M\"unchen, Germany}   and A. Burkert$^{1}$\thanks{E-mail: burkert@usm.uni-muenchen.de}\\
$^{1}$Universit\"atssternwarte, Scheinerstr. 1, 81679 M\"unchen, Germany,\\
$^{2}$Insititute of Astronomy, Madingley Road, Cambrige, CB3 0HA, UK}
\begin{document}

\date{Submitted to MNRAS ------; Accepted ------------}

\pagerange{\pageref{firstpage}--\pageref{lastpage}} \pubyear{2004}

\maketitle

\label{firstpage}

\begin{abstract}
We present a detailed investigation of the relation between
the orbital content of merger remnants and observable properties of
elliptical and S0 galaxies. Our analysis is based on the 
statistical sample of collisionless  mergers of disk galaxies with
different mass ratios and orbital parameters, published by Naab \& Burkert. We use the spectral 
method by Carpintero \& Aguilar to determine the orbital content of every remnant and correlate it 
with its intrinsic shape, and its projected kinematic and photometric properties. We discuss the influence 
of the bulge component and varying pericenter distances. The classified orbit families are box orbits, minor 
axis tubes, inner and outer major axis tubes, and boxlets. In general, box orbits dominate the inner parts 
of the remnant. Major and minor axis tubes become dominant at intermediate radii and boxlets at large radii. 
The two most abundant orbit classes are the minor axis tubes and the box orbits. Their ratio seems to 
determine the basic properties of a remnant. On average, the fraction of minor axis tubes increases by a 
factor of two from a merger mass ratio of 1:1 to 4:1, whereas the fraction of box orbits decreases by 10\%. 
At a given mass the central velocity dispersion of a remnant scales with the ratio of minor axis tubes to box 
orbits. Interestingly, the division line between rotational supported systems and pressure supported objects, 
$(v_{maj}/\sigma_0)^*=0.7$, turns out to coincide with a box to minor axis tube ratio of unity. 
The observed $h_3$-$v/\sigma$ anti-correlation for ellipticals can not be reproduced by collisionless 
merger remnants. We propose that this can only be reconciled by an additional physical process that 
significantly reduces the box orbit content. Remnants which are dominated by minor axis tube orbits have 
predominantly disky projections. Boxy remnants have always a box to minor axis tube ratio larger than one. 
This study will enable to identify observed ellipticals that could have formed, in the collisionless limit, 
by gas-poor disk mergers. In addition, it demonstrates how observable properties of spheroidal stellar systems 
are connected with their intrinsic orbital structure.

\end{abstract}


\section{Introduction}
Elliptical galaxies show a richness of distinct substructure in their isophotal 
shape and velocity anisotropy. \citet{Bender_88} and \citet{BDM_88} found that
elliptical galaxies can be divided into objects with boxy and
disky isophotal shape. Interestingly, ellipticals with boxy isophotes show X-ray emission 
in excess of discrete sources indicative of a hot gaseous halo, rotate slowly, 
and have higher luminosity while disky ellipticals are fast rotators, with low luminosities
and no detectable X-ray halos \citet{Bender_89}. These features are probably linked
to the formation process of these galaxies. \citet{TT_72} were the first to propose that
ellipticals could originate from mergers of two disk galaxies. 
Subsequent N-body simulations of collisions of disk galaxies showed that merger remnants 
mimic important features of elliptical galaxies like global kinematic and photometric properties, 
kinematic misalignments or kinematically decoupled cores 
(\citealp{Barnes_92}, \citealp{HERN92}, \citealp{HERN93}, \citealp{Heyl_94}, \citealp{Heyl_96}, 
\citealp{Weil_96}, \citealp{NBH}, \citealp{BB_00},\citealp{Cretton_01}, \citealp{NB03}, \citealp{BN04}). 
\citet{BSG94} showed that the asymmetry of the line-of-sight-velocity-distribution (LOSVD), correlates 
with the ratio of local rotation velocity to velocity dispersion ($v/\sigma)$. The LOSVD in general 
shows a steep leading wing. These findings have been confirmed by newer observations 
(\citealp{HALL01}, \citealp{PINK03}).    

If we consider ellipticals as pure stellar systems all kinematic and photometric
properties will originate from the projected superposition of the orbits of 
individual stars, which build up the galaxy. Unfortunately, the trajectories of the stars 
themselves are not observable. Reliable constraints about the intrinsic shape can only be 
inferred statistically if a large sample of observed ellipticities is available like for the 
SDSS \citep{Alam_02}. To deproject an individual galaxy and determine its orbital content is 
much harder. The only tool available are Schwarzschild models \citep{SWS79}, 
where a library of orbits is fitted to the light distribution and, if available, to
the detailed kinematics of a galaxy (see \citealp{Rix_97} and references therein). One 
limitation is that most of these models assume that the fitted galaxy is intrinsically 
axisymmetric which might not always be the case in nature \citep{STAT04}.

Simulated galaxies have the advantage that intrinsic and projected properties can 
be studied at the same time. Several authors investigated the orbital content of 
simulated merger remnants. One of the basic results was that the centre of the remnants is dominated 
by box orbits, while minor axis tubes dominate at larger radii \citep{Barnes_92}. 
In agreement with theory they found that major axis tubes dominate in prolate 
remnants. The presence of gas in mergers has a significant impact
on the orbital structure, as the box orbits at the centre are destroyed and minor axis tubes
get more populated \citep{BH_96}. \citet{BB_00} extracted the LOSVD for single orbit classes for
a few merger remnants. Although these studies highlighted the importance of orbital structure and 
sometimes also discussed the effect of viewing angles (\citealp{Heyl_94}, \citealp{Heyl_95}), they were, however,
limited to a small number of merger remnants.

\citet{BS82} originally proposed to use spectral dynamics for stellar dynamical problems. 
\citet{LASK93} devised an algorithm to analyze dynamical spectra with high accuracy, which was also
applied to analytical models of galactic systems (\citealp{PL96};\citealp{PL98}). 
\citet{CA_98} (henceforth CA98) created a fully automated code to classify orbits in arbitrary
two or three-dimensional potentials through spectral dynamics and which we will use in 
all our analysis presented in this paper..
Whereas the orbital classification algorithms used for simulations so far picked up the different orbit families by
checking if they change the sign of the angular momentum or not, the code of CA98 allows to
distinguish orbit classes, which could not have been classified with simpler methods.\\ 
The purpose of this paper is to identify the orbital content of the merger remnants and
connect it with their observable and intrinsic properties. The main ingredients are
a large sample of collisionless merger remnants, software which mimics real life observations 
\citep{NB03} and an orbit classification routine based on CA98.
We study the orbit classes or superpositions of orbit classes, which are responsible for 
photometric properties, like isophotal shape and ellipticity, and kinematic properties of the remnants.\\
A brief introduction to the classification technique is outlined in Section
\ref{sec:class}. In Section \ref{sec:sample} we explain the numerical tests and 
the merger remnants that are included in the sample. The following sections are concerned with global
orbit abundance and radial distribution of orbit classes: Section \ref{sec:results} and section \ref{sec:shape}
describe the relation between the orbital content and intrinsic shape, 
Section \ref{sec:photo} links photometric properties to the orbital structure.
Kinematic properties are discussed in Section \ref{sec:kin}. Finally, we summarize and 
discuss the results in Section \ref{sec:end}.

\section[]{Orbit Classification}
\label{sec:class}
\citet{CA_98} devised a code which automatically calculates the Fourier spectrum 
of the motion of a particle. It finds the most important resonances between the motions 
in x, y and z-direction, which are identified with long, intermediate and short axis of the potential the orbits
are integrated in. We speak of a resonance when
\begin{equation}
l\omega_1 +m\omega_2+n\omega_3=0
\end{equation}
for a non-trivial combination of integers l, m and n. Here
$\omega_{i, i=1,2,3}$  denote the leading frequency in the direction of
motion (in our case Cartesian coordinates). Every line in a 
spectrum of a 3-dimensional orbit can be expressed by a linear combination of 
of up to three frequencies. These frequencies are also termed {\it fundamental frequencies}.
If the code finds a fourth fundamental frequency the orbit is classified as {\it irregular}. Such
an orbit has only one integral of motion, the energy.

Every orbit is analyzed at least twice, once with the full trajectory and once with some part 
removed. If this procedure results in two different classifications, then a third pass is made.
However, if three different classifications are found then the orbit can not be classified. 
CA98 found very few cases in their analysis
of analytical potentials, where this happened. We do find a low fraction
of orbits (roughly 10\%), which change their family during an integration   
and therefore fail to be assigned to an orbit class. Some loss seems to be inevitable
in noisy N-body potentials. These orbits are termed {\it not classified}.

The resonances are the determining factor for the orbital classes and
we divide the orbits up in five broad classes. The {\it minor axis tubes}   
have a 1:1 resonance of the motion in x and y-direction, or l=1, m=1 and n
arbitrary. Similarly for the {\it inner major axis tube} and {\it outer major axis tube}
classes, but with a 1:1 resonance of motion in the z and y-direction (m=1 and n=1). The code 
distinguishes these two orbit classes by a geometric, rather than spectral criterium: the inner major axis tubes are 
elongated and stay relatively close to the major axis, while the outer major axis tubes extend perpendicular
to it and have a round shape. {\it boxlets} have either one or three resonances, barring a 1:1 resonance, 
because then they would have a definite sense of rotation. For example, a l:m:n=4:3:1 resonance 
would classify as a boxlet. The {\it box orbits} are the only orbit class which have no resonance.

The code was tested by CA98 on the singular, triaxial logarithmic potential, first 
analyzed by \citet{SWS93}. They found a very good agreement between their 
spectral method and the much more time consuming investigation of the surfaces of section. 
For mathematical details on the calculation of the frequencies  we refer to the paper of CA98.   

There are several features of the method by CA98 which makes it attractive for
use on N-Body systems: 1.) It can distinguish more orbit classes than
if we would just check if the orbits change the sign of a component
of their angular momentum. 2.) It has an inbuilt routine, which checks
the geometric differences of inner and outer major axis tubes, which
could otherwise not be identified by the spectral method. 3.) It
corrects automatically for the center of the potential. 4.) Slight twists of
the principal axes can be taken into account  by the code, as long as the opening
of a tube orbit is large enough, i.e. the orbit has a high angular momentum.
\begin{table}
\begin{center}
\caption{Number of particles in merger remnants with 
different mass ratios \label{tab:num}}
\begin{tabular}{|l|c|c|c|c|c|c|}
\hline \hline
Mass    & $N_{lum}$& $N_{dark}$     \\
Ratio   &          &             \\
\hline
1:1     & 160000   &  240000      \\
2:1     & 120000   &  180000      \\
3:1     & 106666   &  160000      \\
4:1     & 100000   &  150000      \\
\hline 
\end{tabular}
\end{center}
\end{table}

\section{Remnant Selection and Numerical Tests.}
\label{sec:sample}
\subsection{Figure rotation}
The statistical set of 112 merger simulations of disk galaxies with mass
ratios of 1:1, 2:1, 3:1 and 4:1 is described in detail in
\citet{NB03}. Here we only give a brief summary.
We used the following system of units: gravitational constant G=1,
exponential scale length of the larger disk $h=1$ and mass of the
larger disk $M_d=1$. Each galaxy consists of an exponential disk, a
spherical, non-rotating bulge with mass $M_b = 1/3$, a Hernquist
density profile \citep{HERN90} with a scale length $r_b=0.2h$  and a
spherical pseudo-isothermal halo with a mass $M_d=5.8$, cut-off radius
$r_c=10h$ and core radius $\gamma=1h$. The equal-mass mergers were
calculated adopting in total 400000 particles with each galaxy
consisting of 20000 bulge particles, 60000 disk particles, and 120000
halo particles. The low-mass companion contained a fraction of $1 /
\eta $ the mass and the number of particles in each component with a
disk scale length of $h=\sqrt{1/\eta }$, as expected from the
Tully-Fisher relation \citep{PT92}. The galaxies
approached each other on nearly parabolic orbits with an initial
separation of $r_{sep} = 30$ length units and a pericenter distance of
$r_p = 2$ length units. 16 different initial disk
inclinations were selected in an unbiased way, following the
procedure described by \citet{Barnes_98}. 

In addition to the already published sample we simulated 16 1:1 merger 
remnants with an initial pericenter distance three times larger than in
the original sample and 16 1:1 mergers with the original pericenter distance, but
without bulges. We will discuss only the general properties of this control 
sample in Section \ref{sec:gen}
  
All simulations have been performed using the newly developed
tree code VINE (Wetzstein, Nelson, Naab \& Burkert, in prep.)
in combination with GRAPE-5 \citep{Kawai}. The Plummer-softening for the 
force calculation was set to $\epsilon = 0.05$. Time integration was 
performed  with a Leap-Frog integrator with a fixed 
time-step of $\Delta t = 0.04$.  

In this paper we restrict the orbit analysis to merger remnants
without significant figure rotation. After the merger of the
central parts of the galaxies was complete we allowed the remnants to
settle into equilibrium for approximately 10 dynamical times (the
half mass rotation period of the more massive progenitor 
disk). Thereafter we computed the amplitude and the phase of the $m=2$
mode for 20 consecutive snapshots separated by 0.05 dynamical times. 
Remnants with phase differences larger than $\Delta \phi=0.2$
are considered to have figure rotation and were removed from the sample. 
In total we excluded 18\% of the 1:1, 12\% of the 2:1, 15\% 3:1 and 18\% 
of the 4:1 remnants from the sample. The classification of orbits in
rotating potentials requires additional care \citep{PF91} and
will be discussed in a forthcoming paper.

\subsection{Potential Reconstruction} 
To calculate the orbits of individual particles we fitted the
potential of every merger remnant using a self-consistent field method
(SCF, \citealp{HO}). With this method the Poisson equation is solved by expanding the potential 
and the density in a bi-orthogonal basis set $[\rho_{nlm}(r),\Phi_{nlm}(r)]$ as
\begin{equation}
\rho(r)=\sum_{nlm} A_{nlm}\rho_{nlm}(r)
\end{equation}
and
\begin{equation}
\Phi(r)=\sum_{nlm} A_{nlm}\Phi_{nlm}(r),
\end{equation}
where n denotes the number of radial expansion terms and l,m the angular terms. 
For more details and tests see \citet{Shun97} and \citet{HH95}. The code used 
in this paper was kindly provided by Shunsuke Hozumi. Two different sets of basis 
functions have been widely used in the literature, one proposed by \citet{CB73} 
and one by \citet{HO}, henceforth CB and HO. The first one is based on the Plummer sphere, 
which has a constant density core and the second one on the Hernquist sphere 
\citep{HERN90}, which has a cuspy $\rho \propto r^{-1}$ core. For every basis set it is, however, 
important to choose the proper scale length to represent the potential of every individual 
merger remnant correctly. We kept the number of expansion terms fixed and match the center 
of the true N-body potential iteratively (by varying the scale length) with the 
potential calculated by the SCF code. We found that the forces for the HO basis set deviate 
more strongly in the center from the true forces than the ones calculated from the CB basis set. The 
merger remnants seem not to reach a high cuspiness, therefore the CB basis set was 
applied throughout this paper.

After the potential was reconstructed, we calculated the trajectories for all luminous particles 
using a high order Runge-Kutta integrator \citep{HAIRER} with automatic step size control. 
Every particle was integrated for approximately 30 dynamical times corresponding to 
$\approx 300$ orbits in the central parts and $\approx 30$ orbits around the effective 
radius of the remnant. All particles representing the luminous component were classified. The 
absolute numbers classified in each remnant are laid out in Table \ref{tab:num}. 

As a large number of angular expansion terms, high $l$, is only suitable for very flattened 
systems like disks, more radial terms than angular terms have been used. The same remnant was 
classified with (n=12, l=6), (n=12, l=4), (n=8, l=6), (n=8, l=4) and (n=6, l=4). Table \ref{tab:expan} 
shows that the results hardly depend on the number of expansion terms. Only the runs with l=6 
seem to have a minor effect on the orbit fractions, as the number of not classified orbits and
irregular orbits increases by a few percent. We decided to use a large number
of radial terms, n=12, and a small number of angular terms l=4. This has the 
advantage of having an acceptable calculation time, while keeping
a sufficient number of expansion terms.

\begin{table*}
\caption{Classification results for a 1:1 merger remnant with different choices 
of expansion coefficients. The orbit fractions vary on the percent level. The 
runs with l=6 are slightly noisier and result in a somewhat higher number
of not classified and irregular orbits. \label{tab:expan}}
\begin{tabular}{|l|c|c|c|c|c|c|c|}
\hline 
(n,l)  & Resonant & Minor     &  Outer   &  Inner    &  Box   & Irregular & Not \\
       &  Box     & Axis Tube &  Major Axis Tube  &  Major Axis Tube   &     &  & Classified \\
\hline \hline  
 12,6 & 0.156    &  0.230    &  0.037   & 0.020    & 0.385  &  0.036   & 0.136    \\   
 12,4 & 0.154 	 &  0.234    &  0.036   & 0.019    & 0.408  &  0.029   & 0.119   \\
 8,6  & 0.150    &  0.226    &  0.035   & 0.017    & 0.396  &  0.036   & 0.140   \\ 
 8,4  & 0.150    &  0.233    &  0.036   & 0.017    & 0.416  &  0.028   & 0.119  \\ 
 6,4  & 0.148    &  0.225    &  0.034   & 0.016    & 0.424  &  0.027   & 0.127  \\ 

\hline 

\end{tabular}
\end{table*}  

\subsection{Time Evolution Test}
\begin{figure}
\vspace{1cm}
\centerline{\includegraphics[angle=0, scale=0.37]{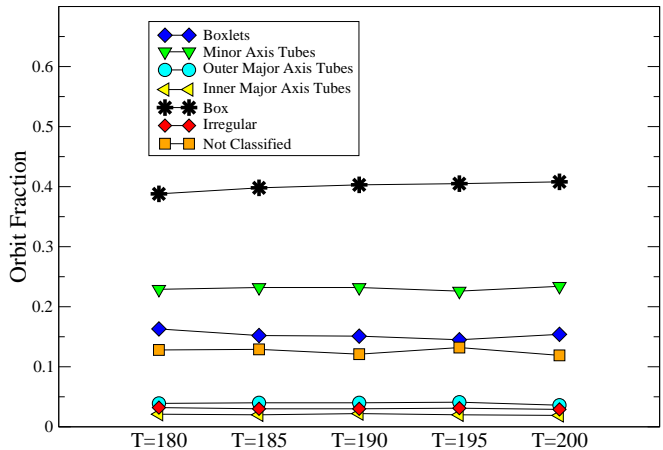}}
\caption{Time evolution of the classification results for a 
  characteristic 1:1 merger remnant over $\approx$ 4 dynamical times. There is no
  significant evolution.}
\label{fig:timevol}
\end{figure}

\citet{NB03} tested that the merger remnants do not change their shape for at 
least 10 dynamical times. It would also be important to know if the orbital content 
is changing or not. In Fig. \ref{fig:timevol} we show five snapshots of a typical 1:1 merger remnant
which are each separated by five computational time units. This is approximately 4 
dynamical times at the half mass radius. All other parameters are kept fix. 
No significant evolution of the orbital content can be observed. 
We therefore can safely use the snapshot at t=200 to analyze the internal
structure of a merger remnant.

\section{Orbits in Merger Remnants}
\label{sec:results}
In this section we discuss the results of the orbit analysis applied
to the sample of collisionless merger remnants. The abundance of
different orbit classes in dependence of the adopted initial conditions,
like mass ratio and impact parameter is highlighted. We examine the radial 
distribution of orbit classes inside a single merger remnant and trace the 
origin of different orbit classes back to the progenitor disk or bulge components. 
 
\subsection{General Results}
\label{sec:gen}
\begin{figure}
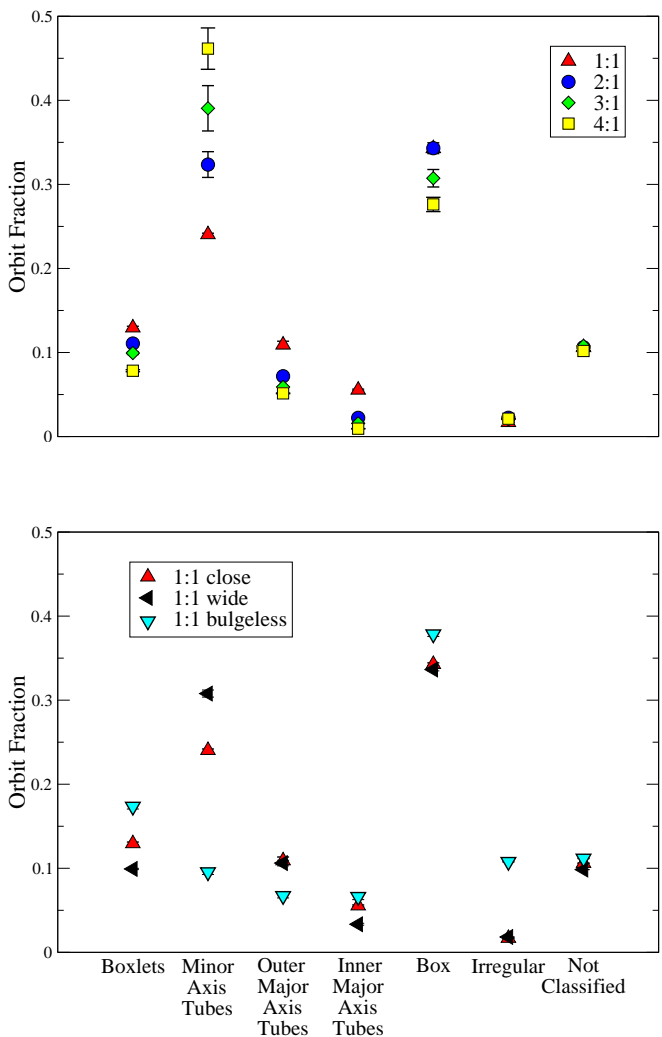

\vspace{1cm}
\centerline{\includegraphics[angle=0,scale=0.37]{mean_orbit.eps}}
\centerline{\includegraphics[angle=0,scale=0.37]{mean_orbit_eqmass.eps}}

\caption{{\bf Top:} Mean orbit fractions for the merger sample 
        with small pericenter distance and different mass ratios. {\bf Bottom:} 
       	Like before, but for equal mass mergers with small and 
	big pericenter distance and without a bulge component}
\label{fig:mean}
\end{figure}

To give an overview over the global orbit abundances we have
determined the particle fraction of a given orbit class for the 40\%
most tightly bound stellar particles of every remnant
which corresponds roughly to the region where global properties
like the anisotropy parameter are observed. Fig. \ref{fig:mean}
shows the result, averaged over all remnants of a given mass
ratio of the progenitor disks.

There are some clear trends connected to the mass ratio of the
progenitors. The number of minor axis tubes is increasing with
increasing mass ratio indicating that in unequal mass mergers it is
more difficult to destroy the orbit population of the progenitor
disks (see \citealp{NB03}). At the same time the boxlet and box orbit
fractions decrease. Major axis tubes of both types only
seem to be populated significantly in equal mass mergers. Over all
there is only a small number of irregular orbits (less than 3\%). The amount 
of not classified, class changing orbits is constant regardless of the mass ratio.
 
The influence of the impact parameter is shown in the bottom panel of Fig.\ref{fig:mean}.
For identical merger symmetries but with larger pericentral distance the minor axis tubes
are populated more strongly than for the close encounter sample. Indeed it is now
very difficult to form prolate remnants. Consequently, the amount of major
axis tubes is much lower. Remnants from bulgeless progenitors show the opposite trend. They have much
less minor axis tubes and many more box orbits of all types than the original sample
with small pericenter distance. This confirms earlier findings that a central mass concentration
tends to destroy the box orbits (\citealp{BH_96}). The amount of irregular orbits 
is higher in the bulgeless remnants due to problems to fit the even shallower potentials with our
basis sets.

\subsection{Radial Distribution} 

\begin{figure*}
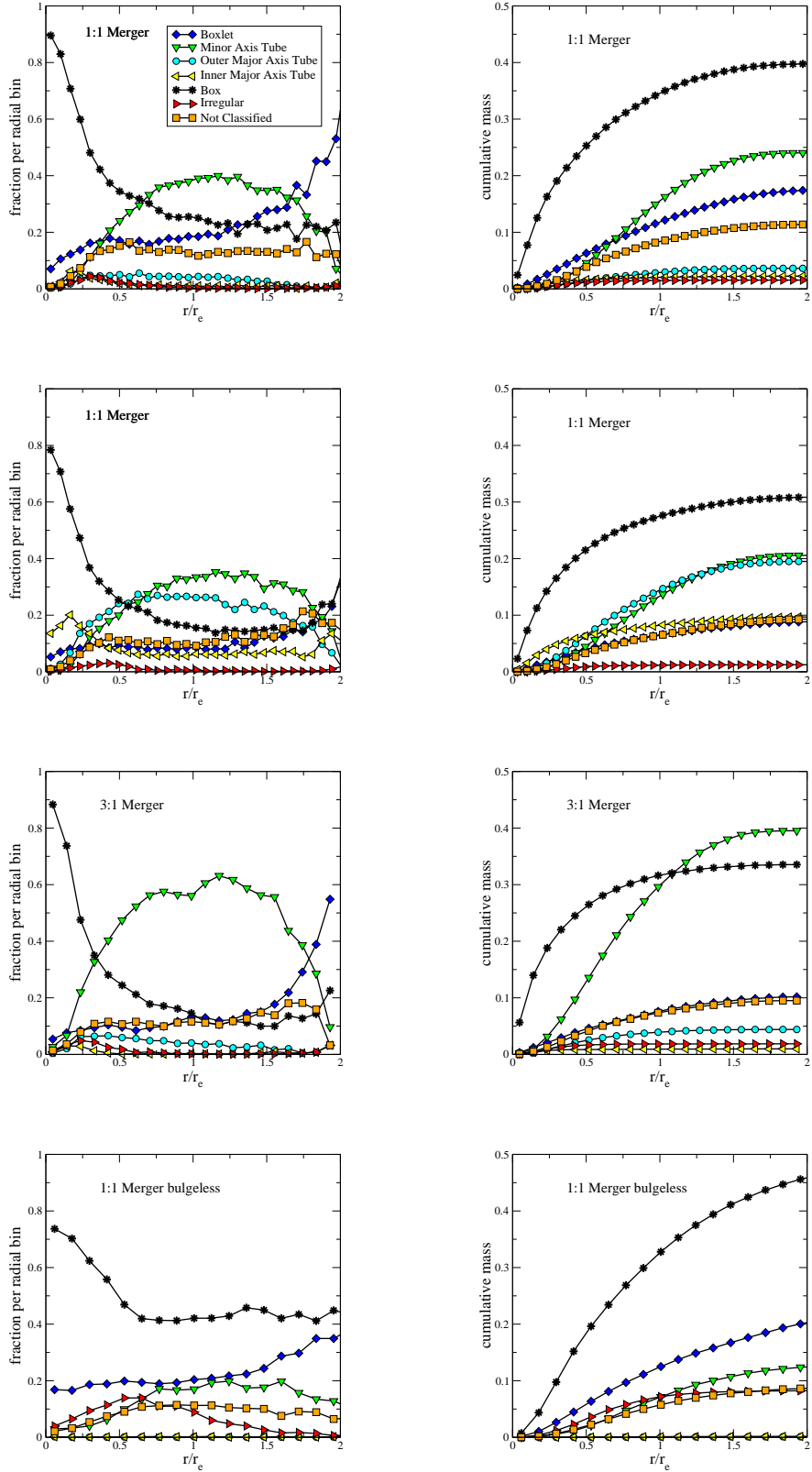

\vspace{.7cm}                                   
\centerline{\includegraphics[angle=0, scale=0.28]{RADIAL_11MCS_5.eps}
\hspace{1.75cm}
\includegraphics[angle=0, scale=0.28]{MASS_11MCS_5.eps}}
\vspace{.85cm}
\centerline{\includegraphics[angle=0, scale=0.28]{RADIAL_11MCS_7.eps}
\hspace{1.75cm}
\includegraphics[angle=0, scale=0.28]{MASS_11MCS_7.eps}}
\vspace{.85cm}
\centerline{\includegraphics[angle=0, scale=0.28]{RADIAL_31MCS_14.eps}
\hspace{1.75cm}
\includegraphics[angle=0, scale=0.28]{MASS_31MCS_14.eps}}
\vspace{.85cm}
\centerline{\includegraphics[angle=0, scale=0.28]{RADIAL_11MCN_1.eps}
\hspace{1.75cm}
\includegraphics[angle=0, scale=0.28]{MASS_11MCN_1.eps}}
\caption{{\bf Left Column:} Radial orbit distribution for different merger symmetries and mass ratios. The top
two mergers are equal mass mergers from progenitors with different disk inclinations, but identical pericenter distance.
Third row: 3:1 merger remnant. Bottom: Merger remnant of two equal mass bulgeless disks. 
{\bf Right Column:} Cumulative mass profiles for the same merger remnants}
\label{fig:radial}
\end{figure*}

The different orbit classes dominate at different radii inside the
merger remnants. To show the radial dependence for individual objects, we plot four 
exemplary merger remnants in Fig.\ref{fig:radial}: two 1:1 merger remnants, a 3:1 remnant 
and a remnant of bulgeless progenitors. Two 1:1 mergers have been chosen to demonstrate that different disk
inclinations can lead to different orbital structure, while pericenter distance and mass ratio are identical.
 In the first three remnants the box orbits are most 
abundant at the center. Boxlets appear at all radii and start to dominate at large radii. This phenomenon has been
found in various analytical potential models (see Lees \& Schwarzschild, 1992 for details). The three tube 
orbit families start to dominate at intermediate radii, due to their non-zero angular momentum. 
In general, the mix of orbit classes is complicated, because the merger remnants do not
have constant triaxiality with radius. Even the most prolate remnant, 
(second row, Fig. \ref{fig:radial}) has a significant amount of minor axis
tubes. In a typical oblate remnant only box orbits and minor axis tubes dominate. 
The bulgeless merger, however, is dominated at all radii by box orbits.

The radial distribution of orbit classes has been theoretically tested for a 
wide range of triaxial St\"ackel models by \citet{AR_94}. The general trends they found 
agree well with the distributions in the merger remnants (compare with Fig.6 and 
Fig.8 of \citealp{AR_94}) 

\subsection{Origin of Orbits from Disk or Bulge of Progenitors}
\begin{figure*}
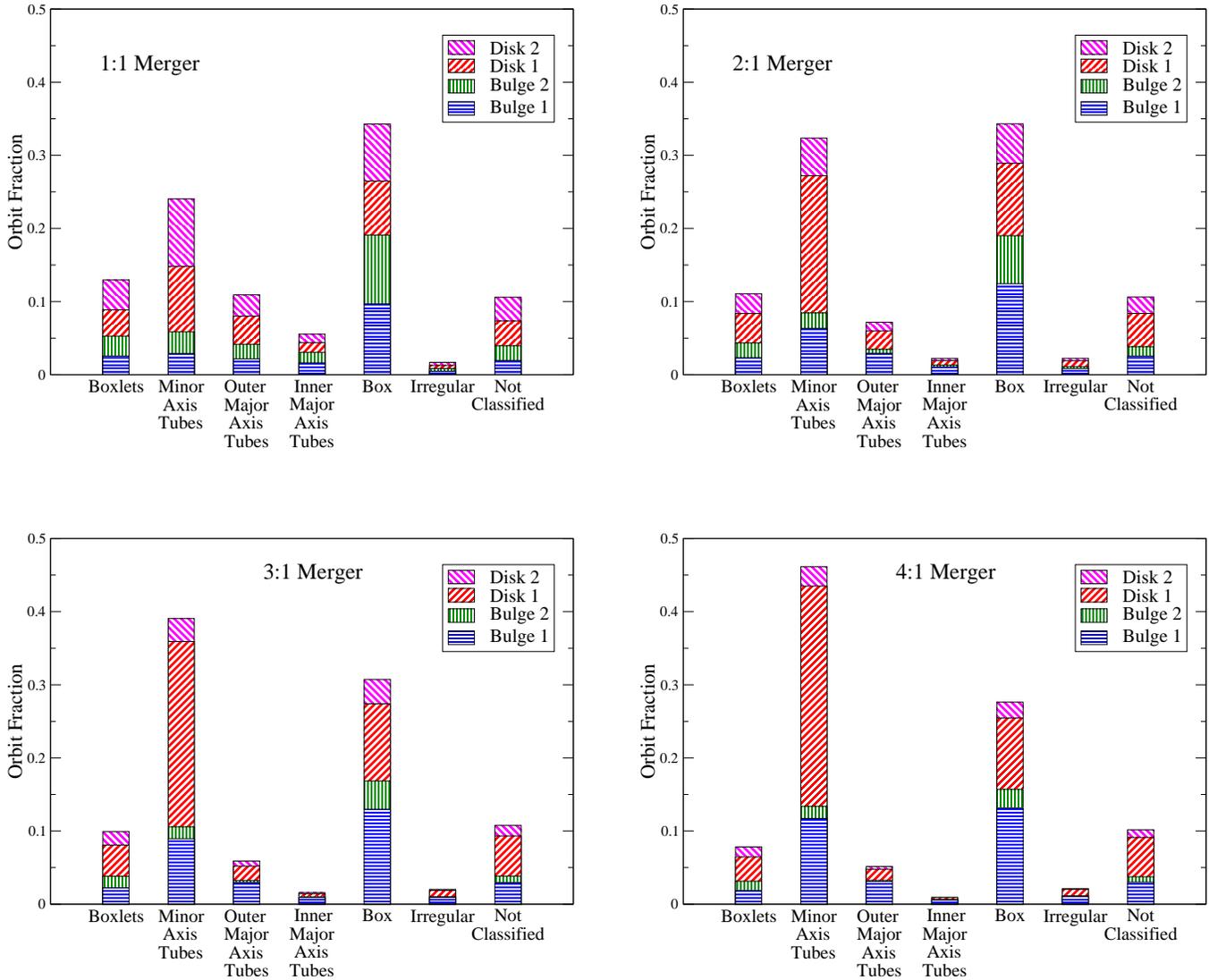

\vspace{1.0cm}
\centerline{\includegraphics[angle=0,scale=0.36]{mean_orbit_orig_11MCS_stack.eps}
\hspace{0.75cm}
\includegraphics[angle=0,scale=0.36]{mean_orbit_orig_21MCS_stack.eps}}
\vspace{1.2cm}
\centerline{\includegraphics[angle=0,scale=0.36]{mean_orbit_orig_31MCS_stack.eps}
\hspace{0.75cm}
\includegraphics[angle=0,scale=0.36]{mean_orbit_orig_41MCS_stack.eps}}
\caption{Origin of the orbit classes from the progenitor bulge or disk components. The histograms are shown
for every mass ratio. The progenitor components denoted with a '1' come from the more massive merging partner 
in the case of unequal mass mergers.}
\label{fig:orig}
\end{figure*}

The merging of two galaxies is a violent and randomizing process. It would be interesting to see
whether certain orbit classes originate from the bulge component or the disk component
of one of the progenitor galaxies. Fig.\ref{fig:orig} shows, averaged over all remnants with
one mass ratio, which fraction of an orbit class originates from the massive disk, massive bulge, light disk and 
light bulge (which of course have the same mass in case of equal mass mergers). 
In the case of the 1:1 mergers we see that in one given orbit class an equal fraction 
originates from one of the two disks or from one of the two bulges. 
Particles originally in a bulge or a disk however populate different orbit classes. The most obvious
trend is that minor axis tubes come mostly from the disk components and box orbits from the bulge component.
Boxlets are slightly more likely to result from one of the disks, as do the outer major axis tubes.

With increasing mass ratio the minor axis tubes become more and more important. In the diagram
for the 4:1 mergers it becomes clear that the overwhelming majority of the minor axis tubes were 
indeed part of the more massive disk. The contribution of massive disk and bulge to the box orbit 
population is remarkably independent of mass ratio.

\section{Orbits and Intrinsic Shape} 
\label{sec:shape}
The intrinsic shape of a triaxial mass distribution is defined by the ratio
of its three principal axes. The principal axes are determined by 
diagonalising the moment of inertia tensor of each merger remnant. The particles are binned 
according to binding energy. That ensures that the subsets of particles follow
the structure of the remnant naturally \citep{Weil_96}. The
triaxiality parameter $T$ is defined as
\begin{equation}
T=\frac{1-(b/a)^2}{1-(c/a)^2},
\end{equation}
where a, b, and c are the long, intermediate and minor axis, respectively.

\begin{figure*}
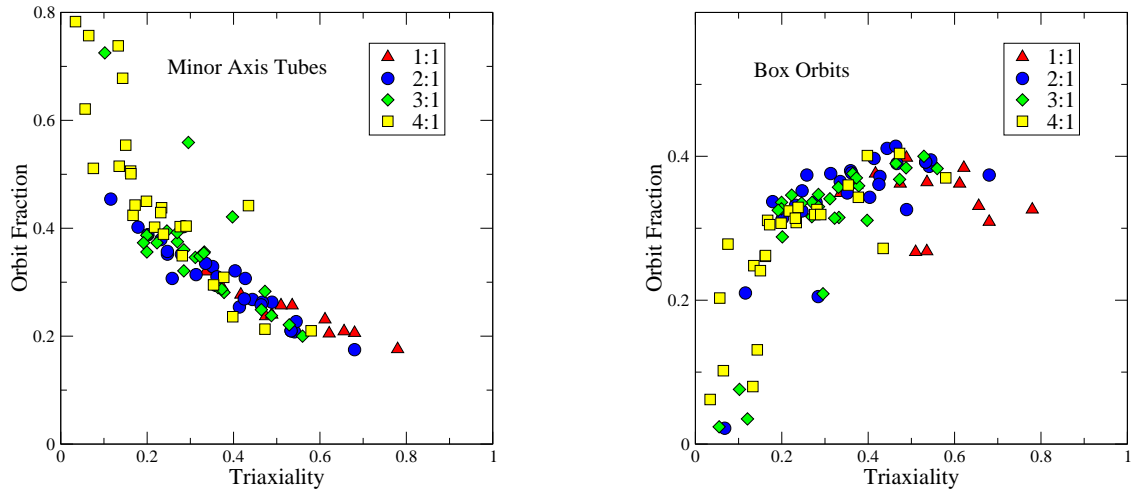

\vspace{1cm}
\centerline{\includegraphics[angle=0,scale=0.38]{correl_trix_minor.eps}
\hspace{1.75cm}
\includegraphics[angle=0,scale=0.38]{correl_trix_box.eps}}
\caption{{\bf Left:} Relation between the triaxiality parameter and the minor 
axis tube fraction. {\bf Right:} Like before, but this time for the box orbit fraction.}
\label{fig:trix}
\end{figure*}

In the following we used the 40\% most bound particles to compute the triaxiality parameter. 
Fig.\ref{fig:trix} shows the distribution of the triaxiality parameter
for the two most important orbit classes: the minor axis tubes and the box orbits.
The minor axis tube fraction rises strongly with decreasing T to 60\% for
oblate shapes ($T=0$) and drops to values as low as 20\%  for very prolate shapes ($T=1$). 
The box orbit distribution peaks at $T=0.5$, i.e. at maximum triaxiality, with a maximum value
of around 40\%. The box orbits depopulate quickly towards the prolate and oblate limits of 
the triaxiality parameter. 

\begin{figure*}
\centerline{\includegraphics[angle=0, scale=0.45]{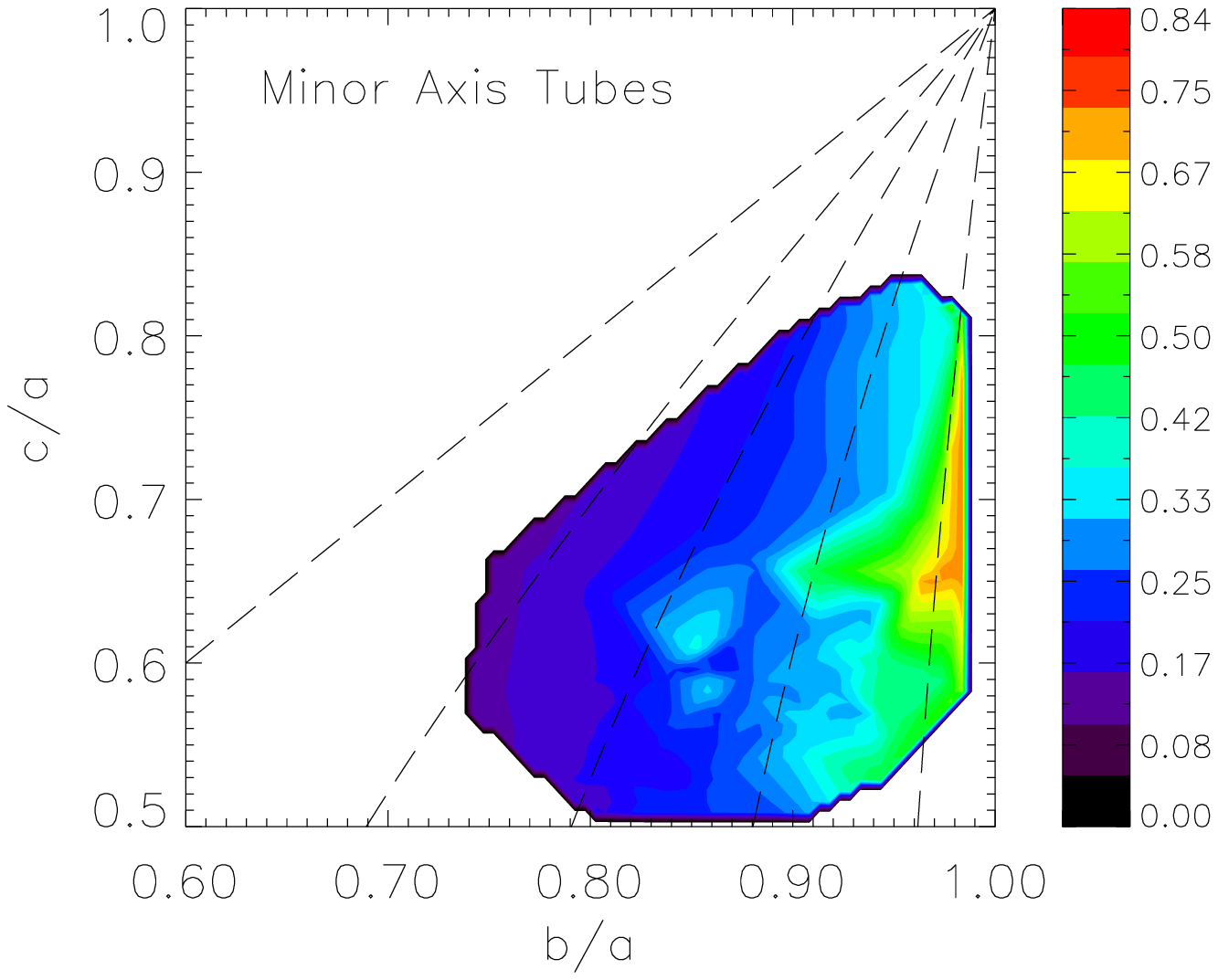}
\includegraphics[angle=0, scale=0.45]{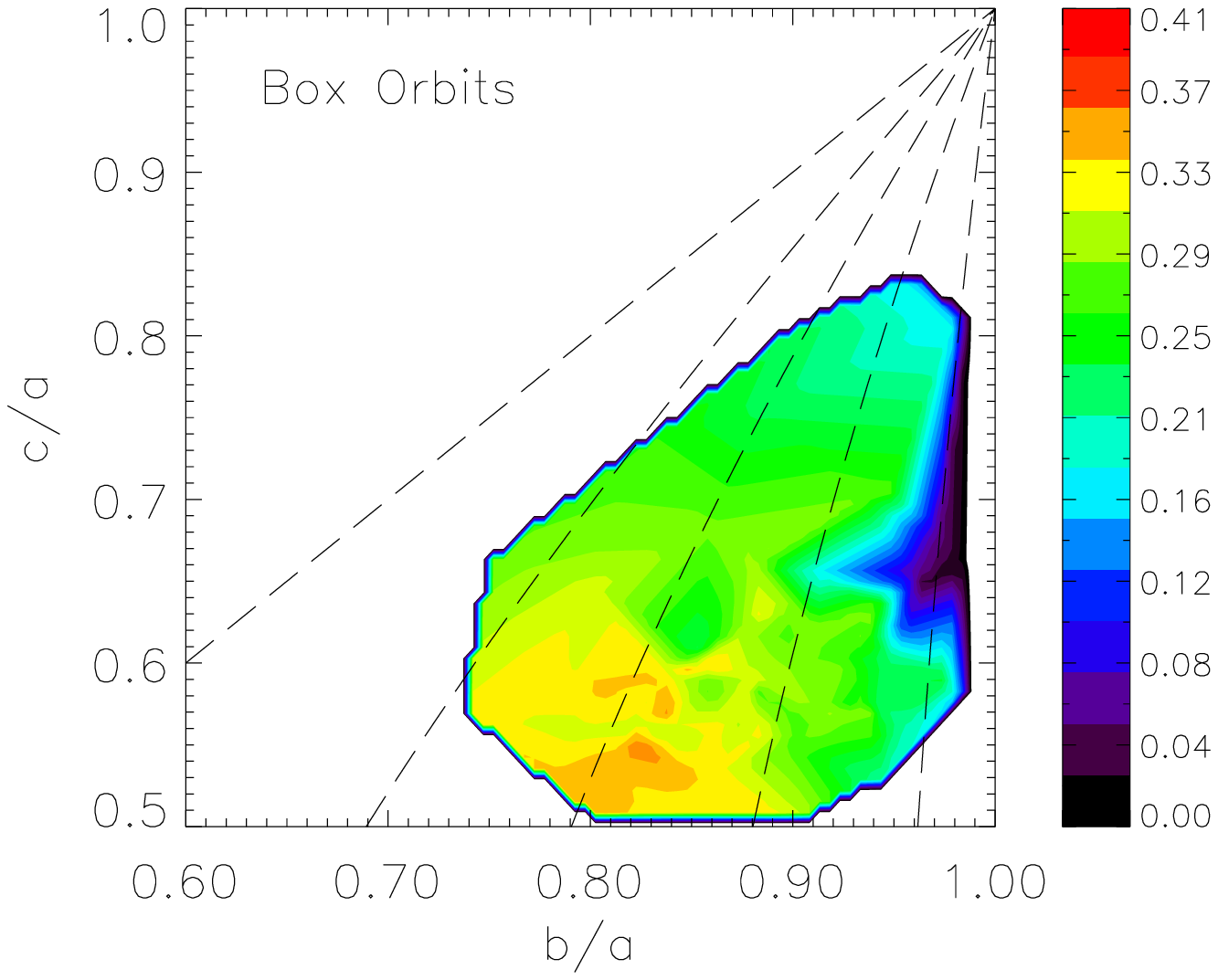}}
\centerline{\includegraphics[angle=0, scale=0.45]{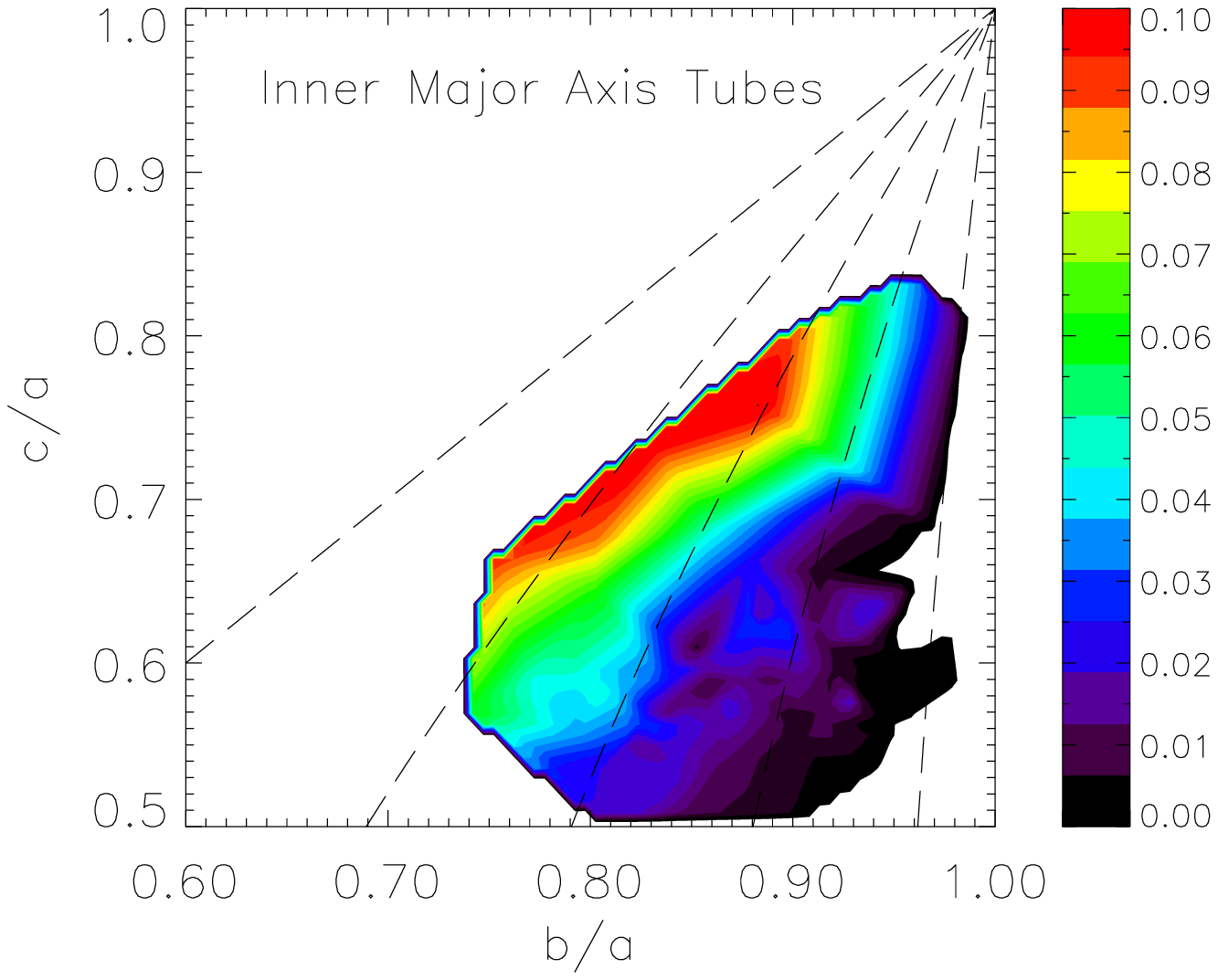}
\includegraphics[angle=0, scale=0.45]{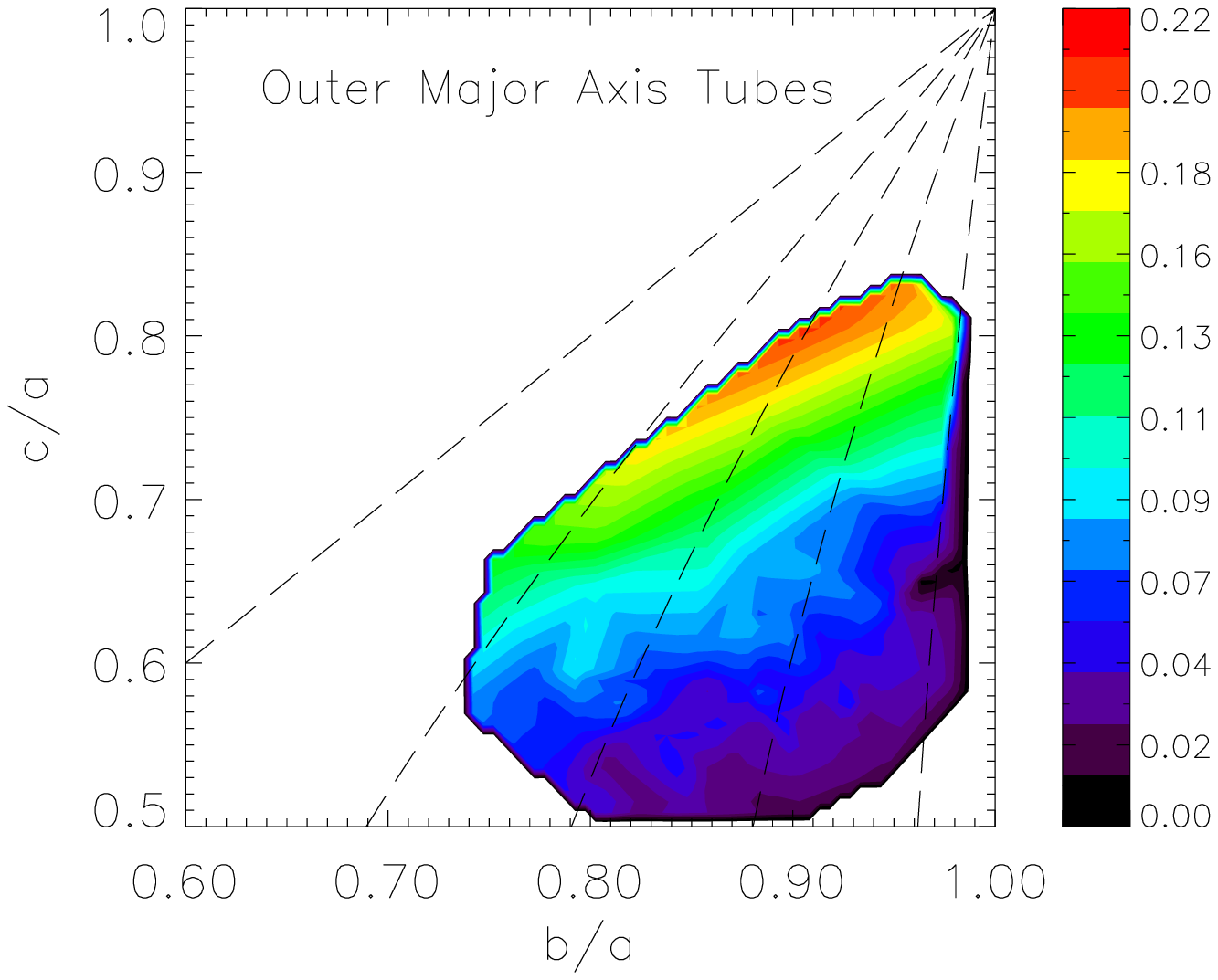}}
\centerline{\includegraphics[angle=0, scale=0.45]{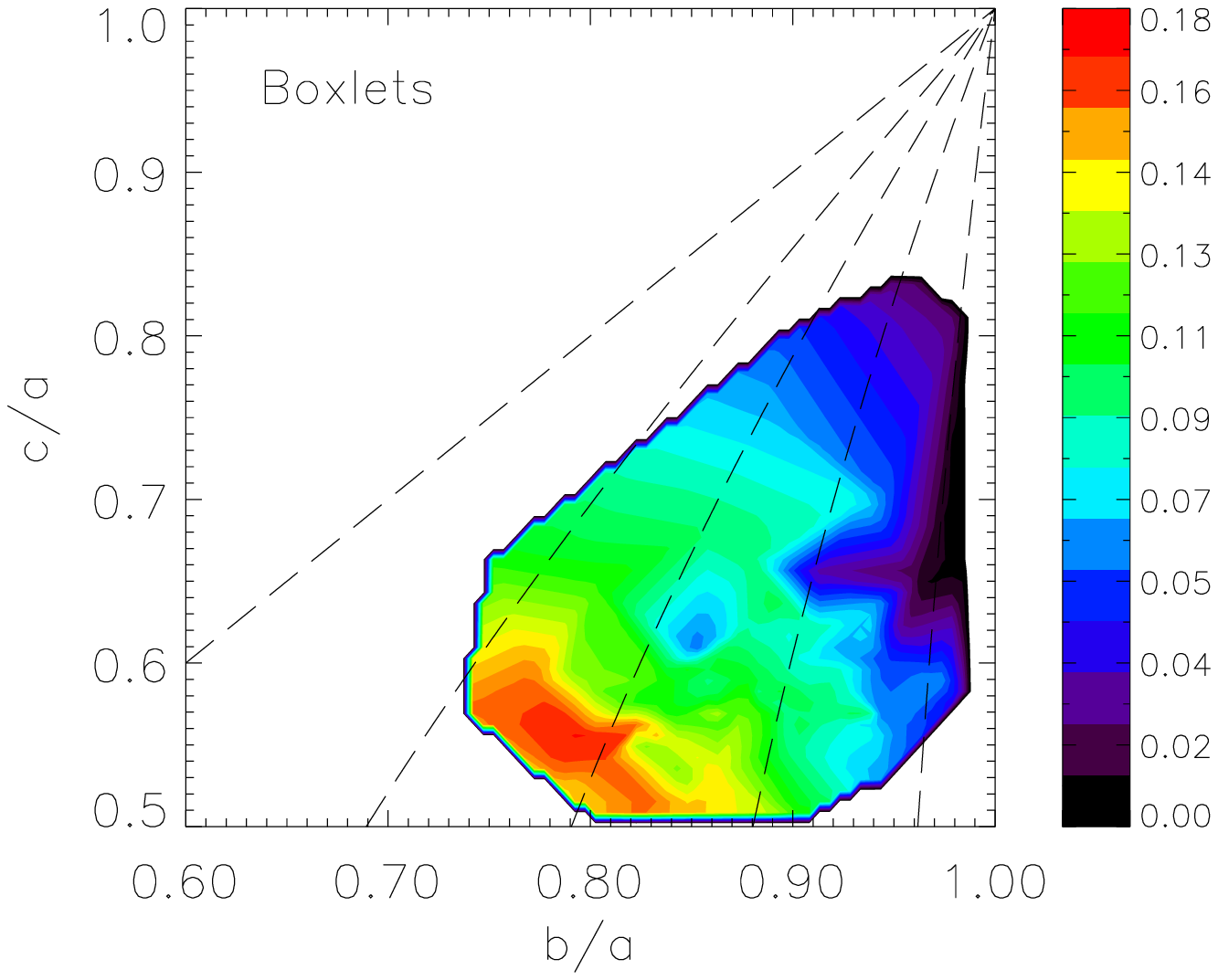}
\includegraphics[angle=0, scale=0.45]{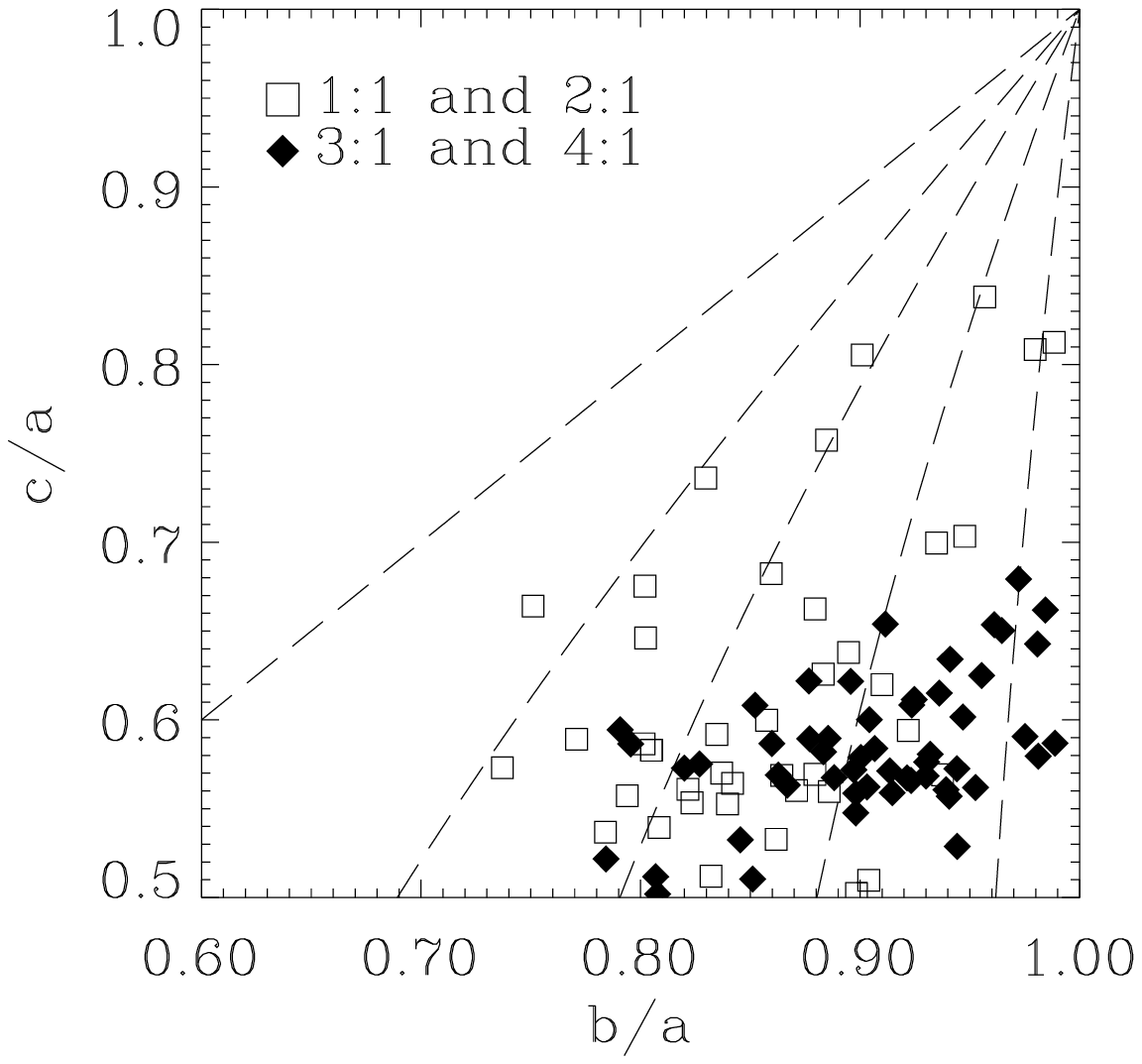}}

\caption{Color coded orbit abundances in axis-ratio space, where red designates the maximum abundance
reached by a given orbit class for all remnants and black the minimum. a, b and c denote 
the long, intermediate and short axis of the $40\%$ most bound
particles of the remnants.  Lines of equal triaxiality are
over plotted. From left to right: T=1, 0.7, 0.5, 0.3. {\bf Top left:}
Minor axis tubes. {\bf Top Right:} Box orbits. {\bf Center Left:}
Inner Major Axis Tubes. {\bf Center Right:} Outer Major Axis
Tubes. {\bf Bottom Left:} Boxlets. {\bf Bottom Right:} Location of
merger remnants in the axis ratio diagram on which the abundances where interpolated} 
\label{fig:axisratio}
\end{figure*}

There is a considerable spread in the results, which in part can 
be explained, when we look at the two-dimensional probability density of the merger remnants 
in axis ratio space (Fig. \ref{fig:axisratio}). 
The same triaxiality can be achieved by different combinations of c/a and b/a. The spread 
in the correlation would be small if the probability contours would follow the lines of equal
triaxiality. This is almost the case for the minor axis tubes. They are most abundant in 
very oblate remnants and depopulate very quickly for other shapes. The upper right panel of
Fig. \ref{fig:axisratio}, which shows the box orbit abundance, the peak is located at 
$T=0.5$. The probability contours are not parallel to the lines of constant triaxiality, 
but also have a gradient along the lines. The abundance of box orbits and minor axis tubes 
is almost anti-correlated. Those two families are the most dominant families in most of the remnants. 
This suggests that many kinematical features might depend on the ratio of the population of those 
two orbit families. We will explore the implications in the following sections.

Outer major axis tubes and inner major axis tubes are not occupying exactly the same positions 
in the diagrams. Inner major axis tubes are abundant in more prolate models, while the 
outer major axis tubes are also common in remnants which tend to the spherical limit (the upper right
corner in the axis ratio plot).
This is the result of the shape of these orbit classes: the inner major axis tubes are
more elongated along the major axis while outer major axis tubes support 
a round and thick shape.

\section{Orbits and Photometric Properties}
\label{sec:photo}
\subsection{Orbital Shapes}
\begin{figure*}
\vspace{1cm}
\centerline{\includegraphics[angle=0,scale=0.38]{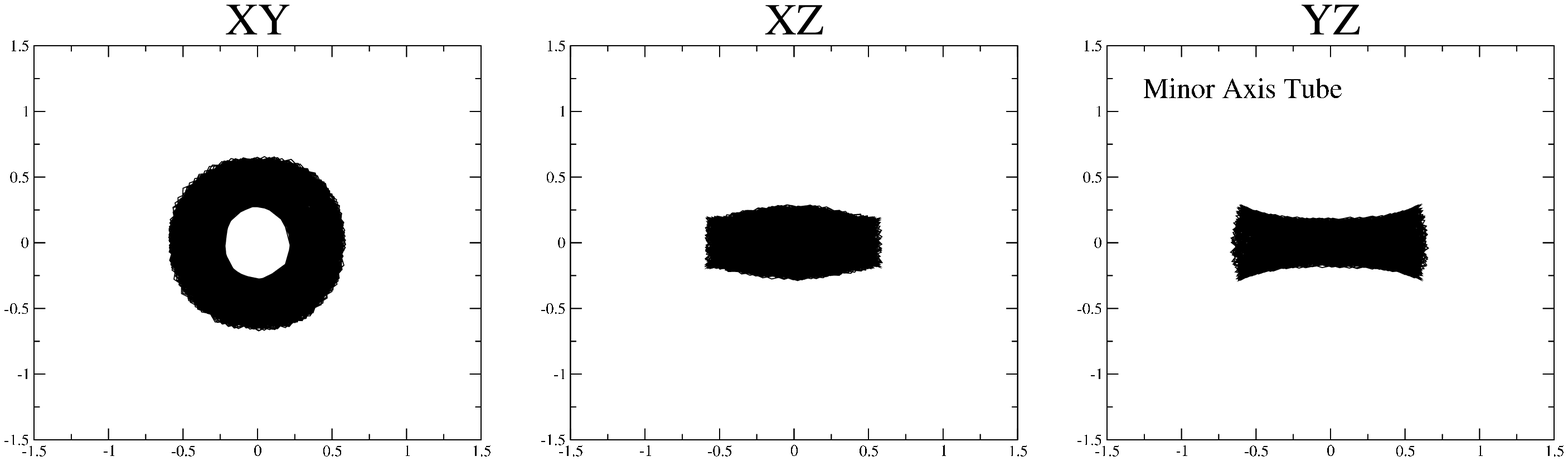}}
\vspace{0.5cm}
\centerline{\includegraphics[angle=0,scale=0.38]{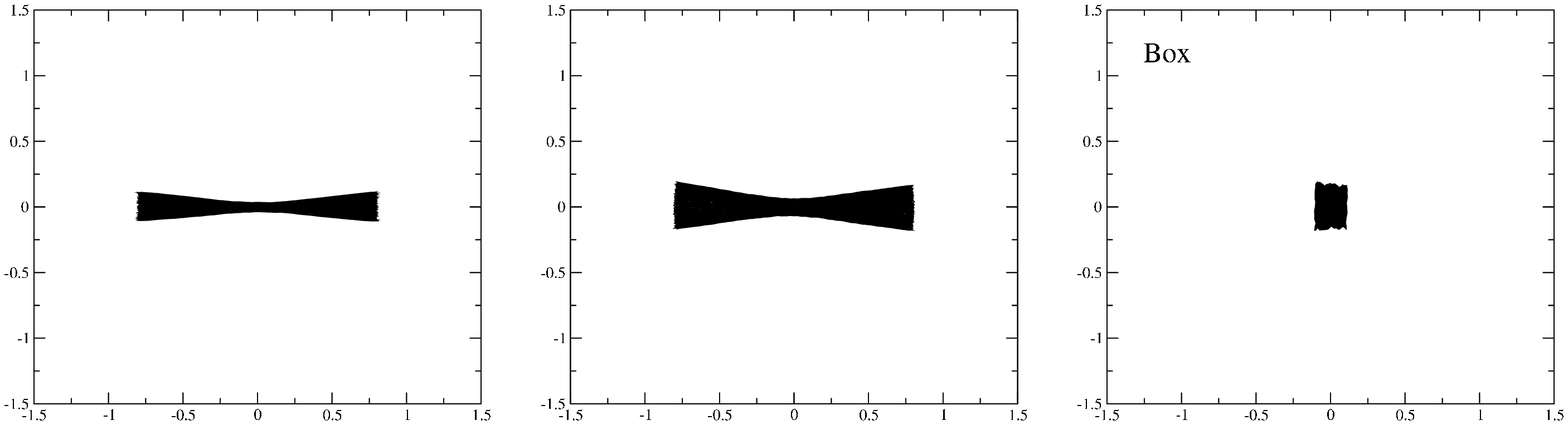}}
\vspace{0.5cm}
\centerline{\includegraphics[angle=0, scale=0.38]{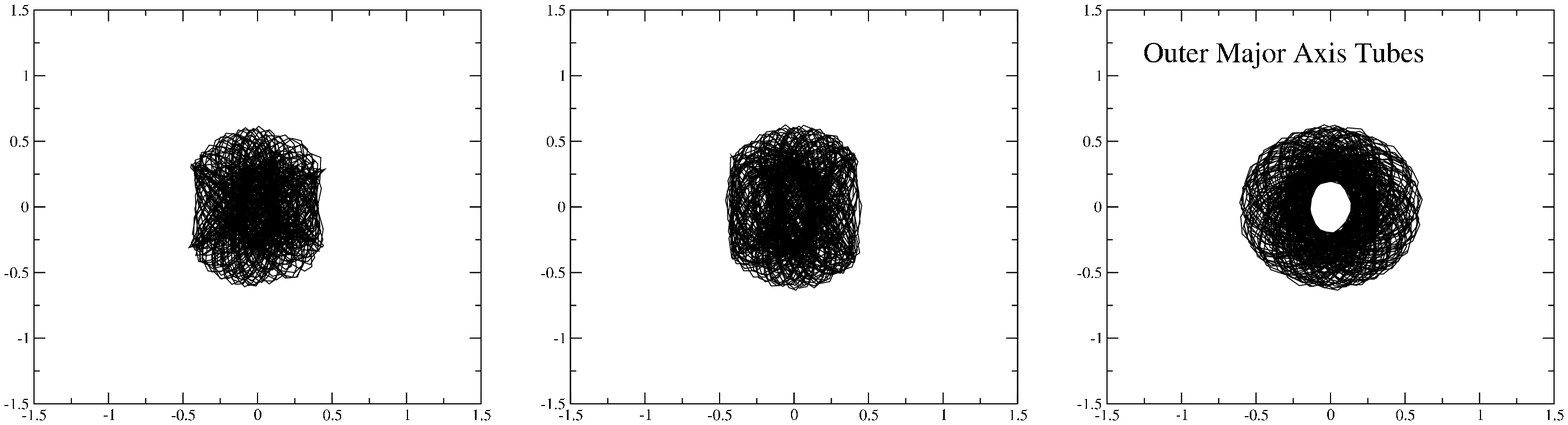}}
\vspace{0.5cm}
\centerline{\includegraphics[angle=0, scale=0.38]{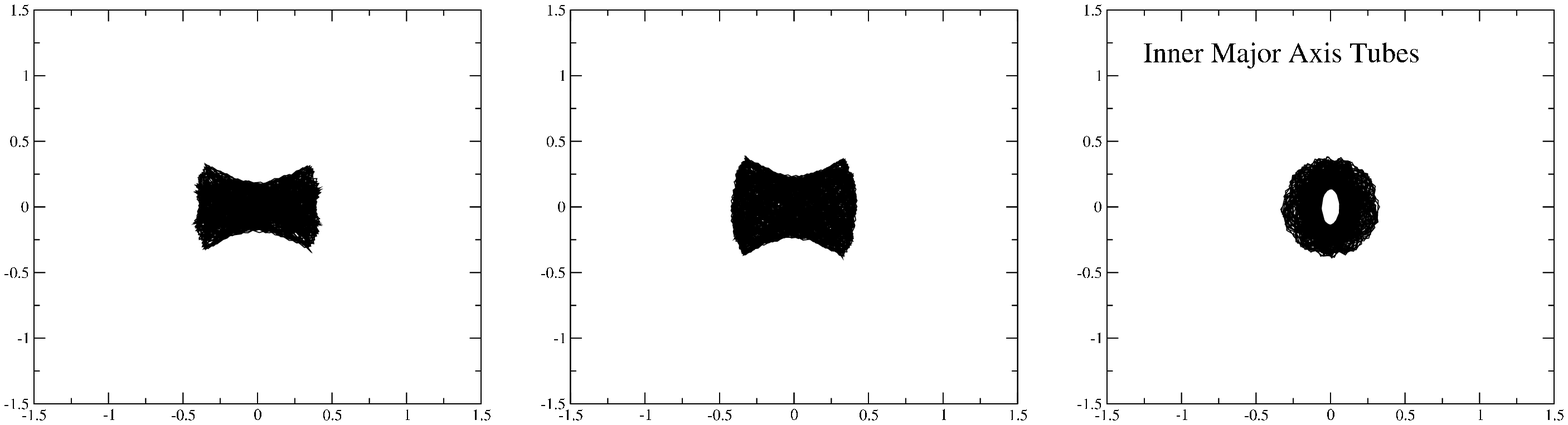}}
\vspace{0.5cm}
\centerline{\includegraphics[angle=0, scale=0.38]{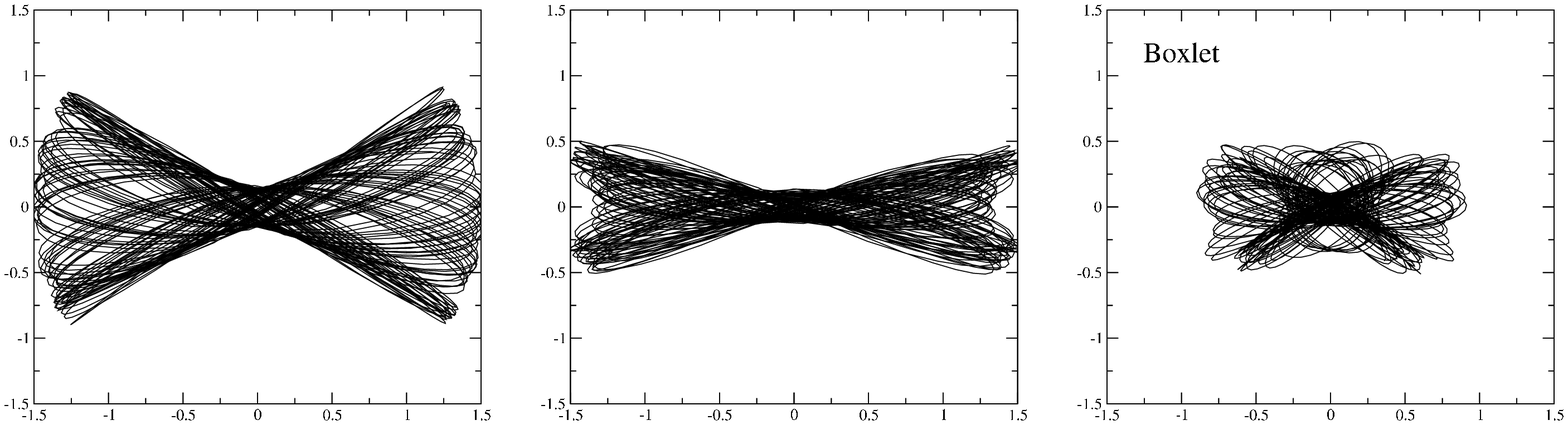}}
\caption{Orbits integrated in the potential of a 1:1 merger. This shows that the projection on the 
three symmetry planes can differ strongly for a given orbit class. Photometric properties change accordingly. 
{\bf From top to bottom}: minor axis tube, box orbit, outer major axis tube, inner major axis tube and boxlet. {\bf From 
left to right}: projection along the short axis (XY), along the intermediate axis (XZ) and along the 
major axis (YZ)}
\label{fig:orbshape}
\end{figure*}

As the remnants deviate significantly from axisymmetry, we have to consider the general case of
orbits in triaxial potentials, when investigating the orbital shapes of individual orbit classes,
e.g. orbit classes which exist in axisymmetric systems like the minor axis tubes, change their appearance in 
triaxial systems. This is illustrated qualitatively in Fig. \ref{fig:orbshape}, where for the five most 
important orbit classes exemplary trajectories are plotted projected onto the three principal planes (XY, XZ, and YZ).   
The projection of the minor axis tube (top row, Fig. \ref{fig:orbshape}) along the short axis appears round and 
has a hole in the center. Particles moving on such an orbit do not come close to the center because of their 
non vanishing in z-direction. The projection along the intermediate axis is flattened and disk-like, almost like one 
would expect in an axisymmetric system. However, the projection along the major axis is slightly peanut-shaped. That might 
imply that minor axis tubes do not contribute to disky isophotes for viewing angles close to 
this projection. The box orbits appear box shaped, with the exception of the projection along
the major axis, where we look end on and the box orbits (second row) do not extend much. The outer and inner major axis tubes
(row three and four) appear round along the major axis projection. The other two projections differ 
between both orbit classes. While inner major axis tubes have an elongated shape along the major axis, the outer 
major axis tubes extend perpendicular to it and in general appear more round. The boxlets (last row) are resonant box 
orbits which have the same general features as their non-resonant brethren, but extend to larger radii. The orbits shown in 
Fig. \ref{fig:orbshape} highlight how the photometric properties of the merger remnants might be influenced in a very 
complicated way by the superposition of different orbit classes as well as by the viewing angle.
Empirically we find that the projections along the principal axes define the boundaries of the photometric properties, 
like ellipticity and isophotal shape. Therefore we focus on those projections in the following.  

\subsection{Ellipticity}
\begin{figure}
\vspace{1cm}
\centerline{\includegraphics[angle=0,scale=0.38]{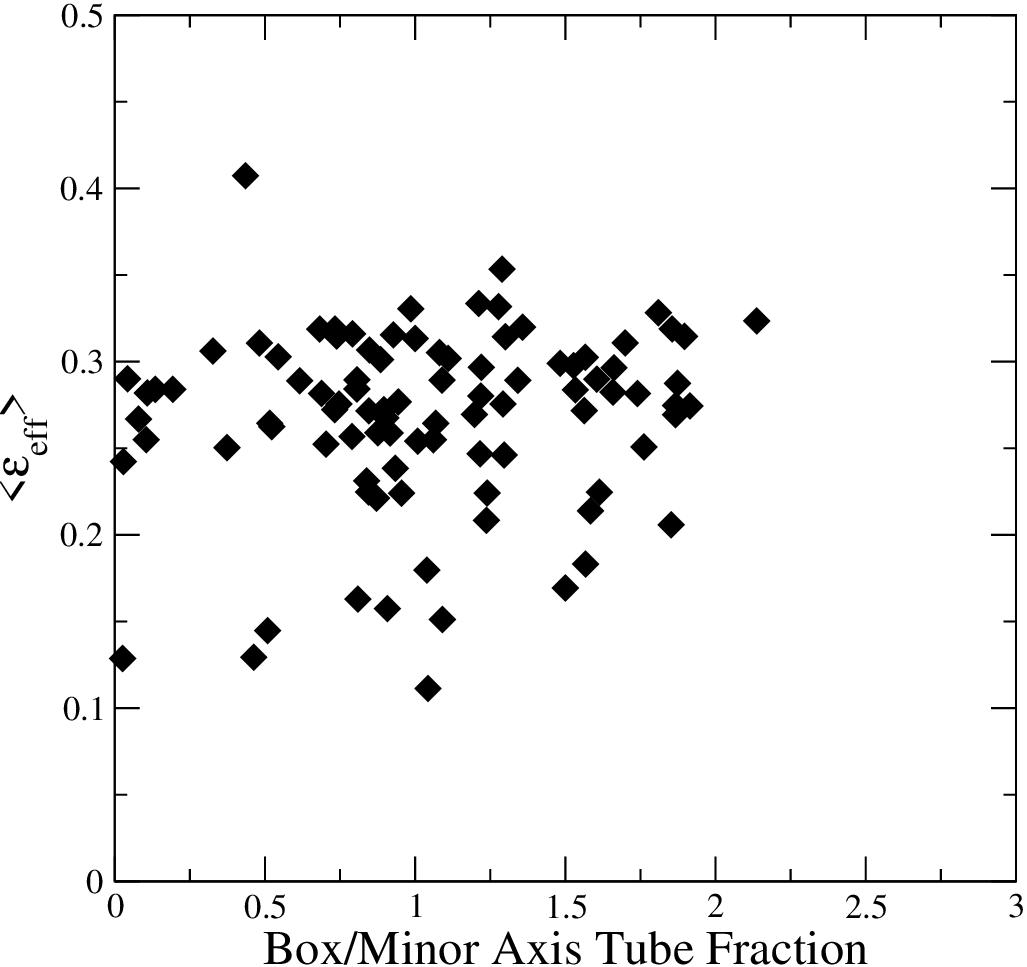}}
\caption{Relation between mean effective ellipticity and the box to minor axis tube ratio.
The mean values lie between 0.1 and 0.3. There is no obvious connection
between the intrinsic structure and the mean ellipticity}
\label{fig:meane}
\end{figure}

\begin{figure*}
\vspace{1cm}
\centerline{\includegraphics[angle=0,scale=0.7]{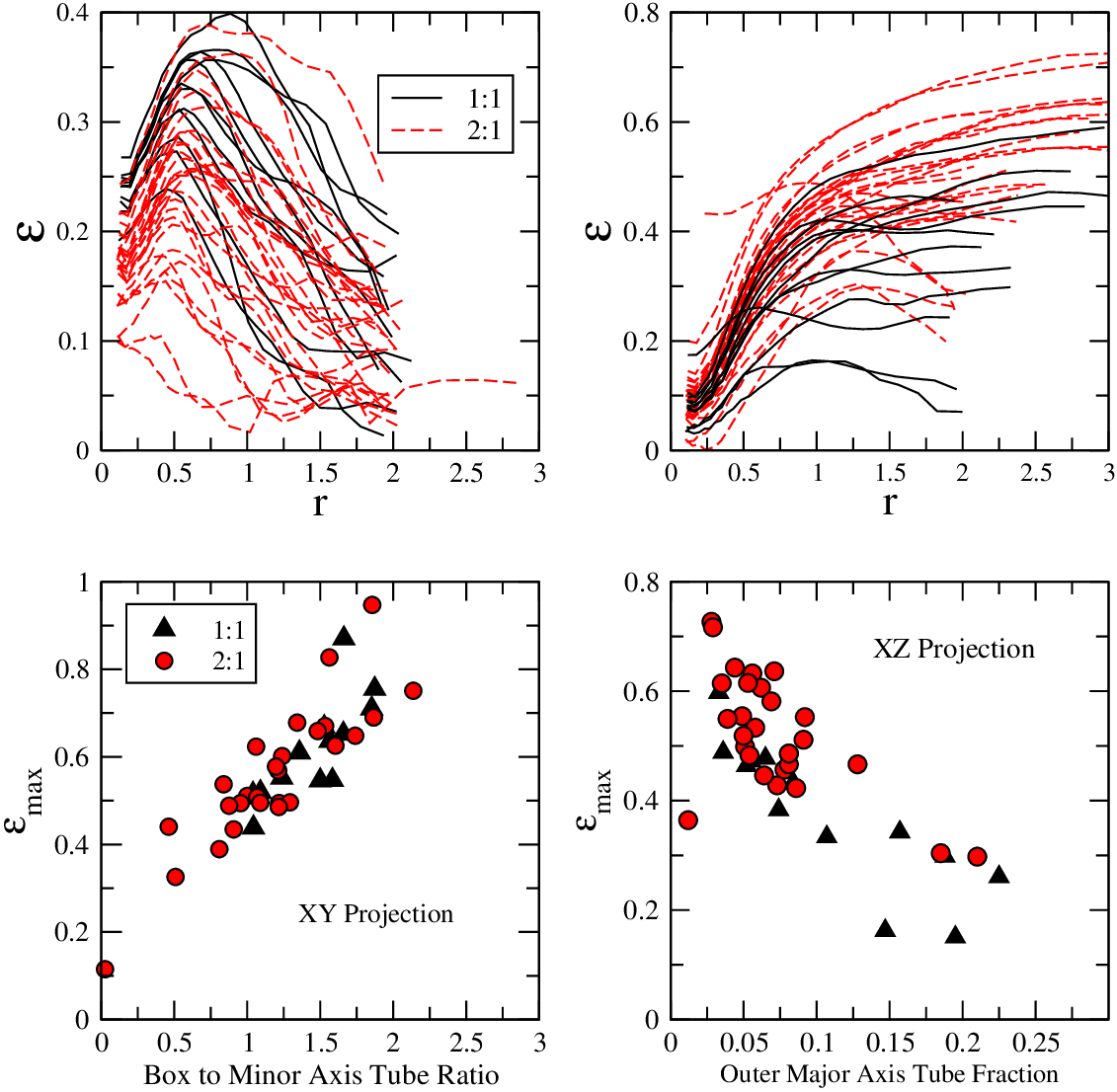}}

\caption{{\bf Top Left:} Radial ellipticity profiles from the projection on 
the XY-plane of 1:1 and 2:1 merger remnants. {\bf Top Right:} Radial ellipticity profiles
from the projection on the XZ-plane of the same merger remnants {\bf Bottom left:} Relation 
between box to minor axis tube ratio  and maximum ellipticity of the XY projection. 
{\bf Bottom Right:} Relation between the outer major axis tube fraction and
the maximum ellipticity of the XZ-plane.}
\label{fig:ellproj}
\end{figure*}

We want to answer the question if the ellipticity is a good indicator of the intrinsic shape and the true 
orbital content. Observationally, the projected ellipticity has to be used to infer statistically 
the intrinsic shape distribution of real elliptical galaxies. In principle we have the advantage to know both the
intrinsic and projected properties, but we defer a detailed analysis to a future paper. 
Instead we simply correlate the average, effective ellipticity, as defined in \citet{NB03}, over fifty projections 
for every remnant with the box to minor axis tube ratio (Fig. \ref{fig:meane}). We see no obvious relation between 
the box to minor axis tube ratio and the mean ellipticity. To investigate the influence of the averaging process,
we examine radial ellipticity profiles for individual galaxies, while keeping the viewing angle fixed.
The radial ellipticity profiles of the XY-projection of the 1:1 and 2:1 remnants 
are shown in Fig. \ref{fig:ellproj}, top left. The shape is similar in each case: 
after a short initial rise the curves peak and then drop to low ellipticities.
There is a trend, that peaks with higher ellipticity are located at larger radii. The value of the 
peak ellipticity correlates nicely with the box to minor axis tube ratio (Fig.\ref{fig:ellproj}, bottom left).
This behaviour becomes clear if we remind ourselves that we superpose for this projection
round tubes with elongated boxes (left most panels of the top two rows, Fig. \ref{fig:orbshape}).
Only a significant fraction of box orbits can lead to an elongated appearance. In the outer
parts, where the minor axis tubes dominate, the remnant appears round. The effect vanishes for 3:1 and 
4:1 remnants, because their box orbit fraction is too low. Ellipticity profiles like this are not 
characteristic for early-type galaxies in general, but individual examples do exist.

The ellipticity profiles for the edge-on projections, e.g. onto
the XZ-plane, first rise steeply and then flatten out to larger radii. 
In Fig. \ref{fig:ellproj}, top right panel, we plot the ellipticity profiles for all 1:1 and 
2:1 merger remnants. Most of them show the described behaviour, but some of them stay at 
remarkably low ellipticities. This time we do not find any correlation with the box or minor axis 
tube fractions. Instead the ellipticity of the edge on projection is very sensible 
to the fraction of outer major axis tubes, as indicated in Fig. \ref{fig:ellproj}, bottom right panel.
Both minor-axis tubes and box orbits populations do not extend very much in the 
z-direction, only the outer major axis tubes do. Even a small fraction of major axis 
tubes seems to be able to lower the ellipticity significantly, i.e. make the shape rounder (see 
right most panels of the upper three rows in Fig. \ref{fig:orbshape}). 
The ellipticity of the 3:1 and 4:1 remnants seems to be determined by minor axis tubes alone. 

\subsection{Mean Isophotal Shape}
\begin{figure}
\vspace{1cm}
\includegraphics[angle=0,scale=0.38]{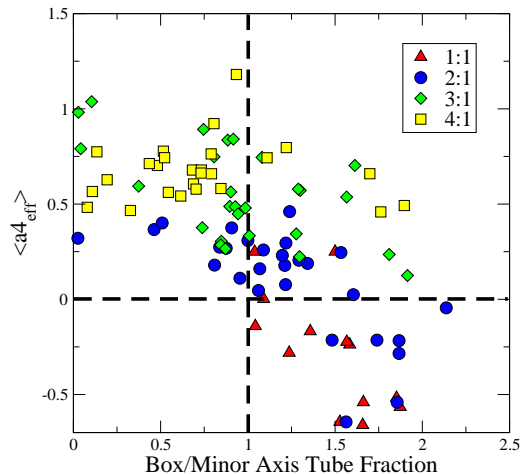}
\caption{Relation between the box to minor axis tube ratio and the effective isophotal shape parameter.
The vertical dashed line divides the box dominated from the minor axis tube dominated merger remnants.
The horizontal line divides the separates the boxy remnants from the disky remnants.}
\label{fig:a4}
\end{figure}

Fig. \ref{fig:a4} summarizes our results for the effective isophotal shape parameter $a4$, averaged over 
50 projections, as defined in \citet{NB03}. For a better discussion we partitioned the plot into four quadrants. 
The bottom right quadrant contains boxy remnants with a box to minor axis tube fraction 
larger than one. Only mergers of the 1:1 and 2:1 fraction can be found here. The opposite 
quadrant (top left) has disky remnants with a dominant minor axis tube 
population and contains no 1:1 mergers. Finally the bottom left quadrant (boxy and dominant
minor axis tubes) has no mergers, while in the upper right one (disky and dominant box orbits)
every mass ratio can be found. The upper right quadrant is difficult to interpret. 
A dependence on the mass fraction can clearly be seen, but on the box to minor axis tube ratio 
not, at least for 3:1 and 4:1 mergers. 

\subsection{Viewing Angle and Isophotal Shape}
\begin{figure*}
\vspace{1cm}
\centerline{\includegraphics[angle=0, scale=0.38]{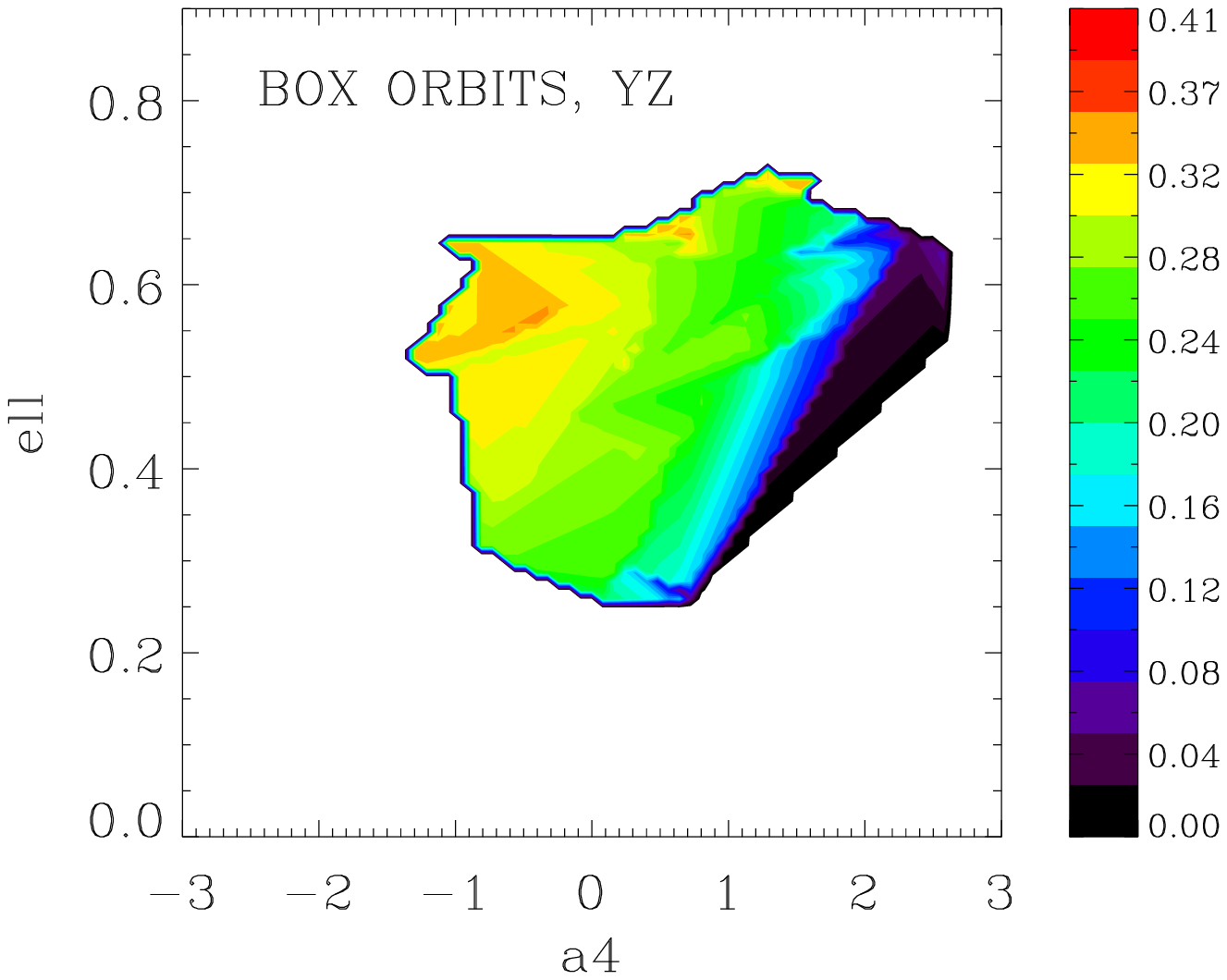}
\includegraphics[angle=0, scale=0.38]{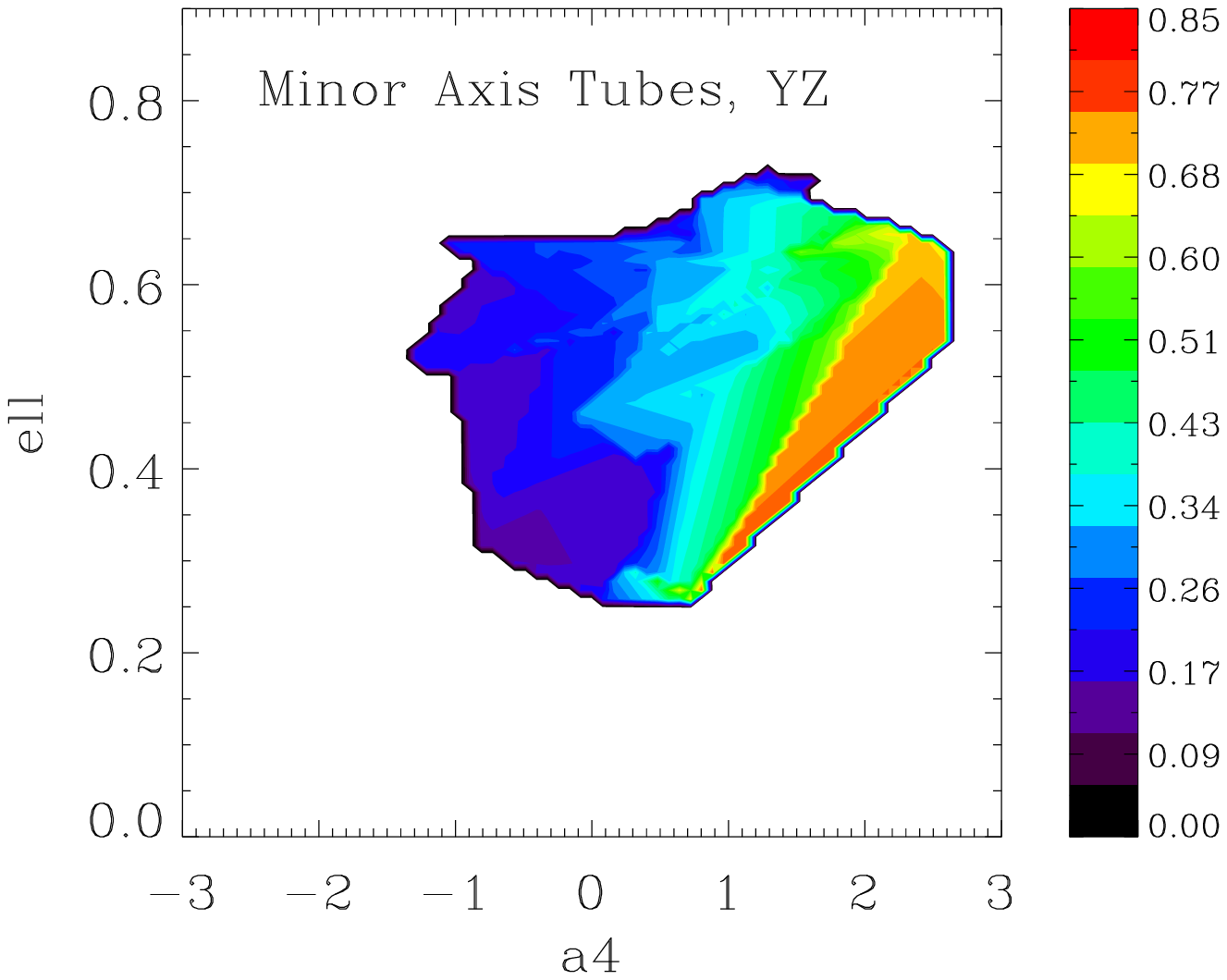}
\includegraphics[angle=0, scale=0.38]{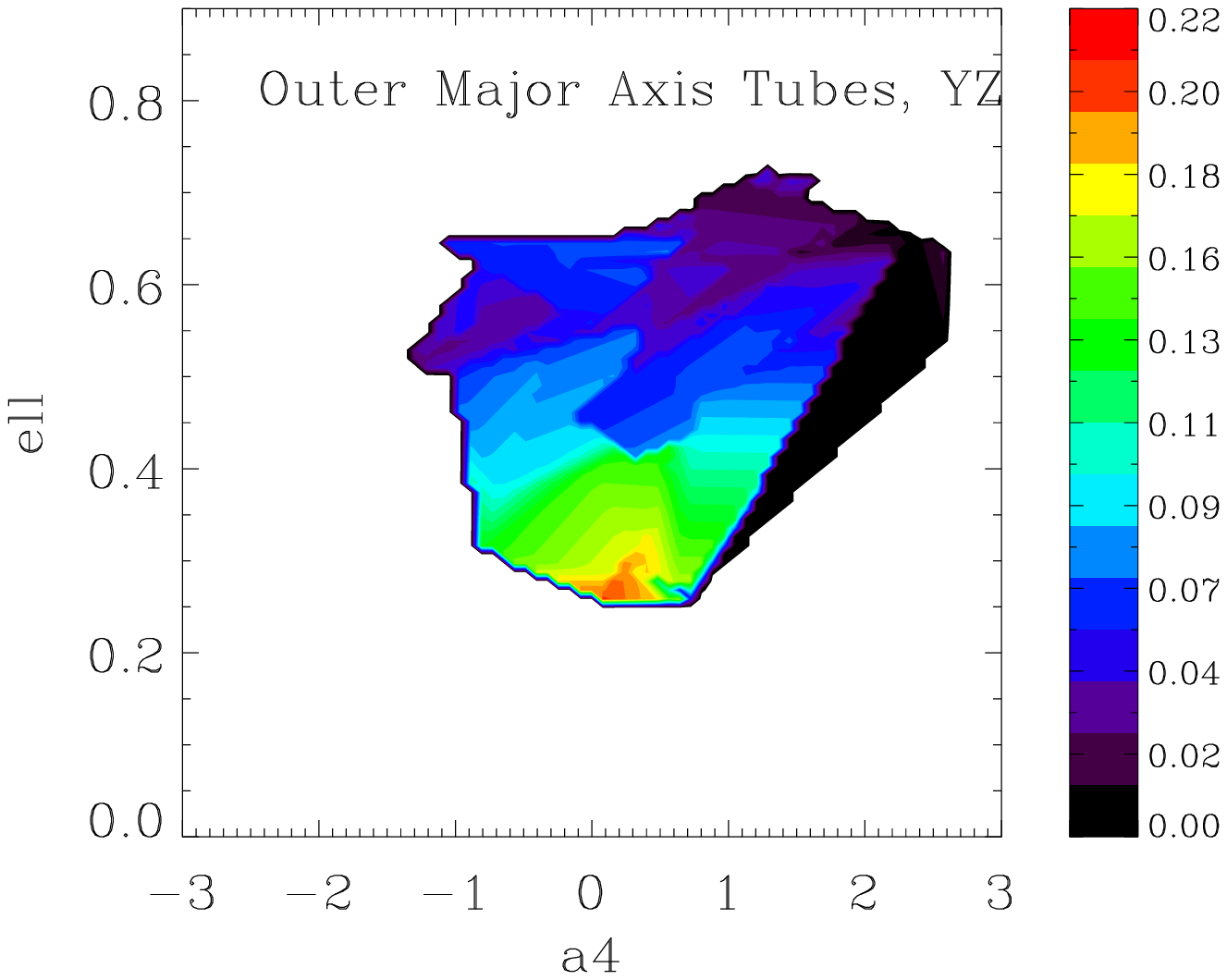}}
\vspace{0.5cm}
\centerline{\includegraphics[angle=0, scale=0.38]{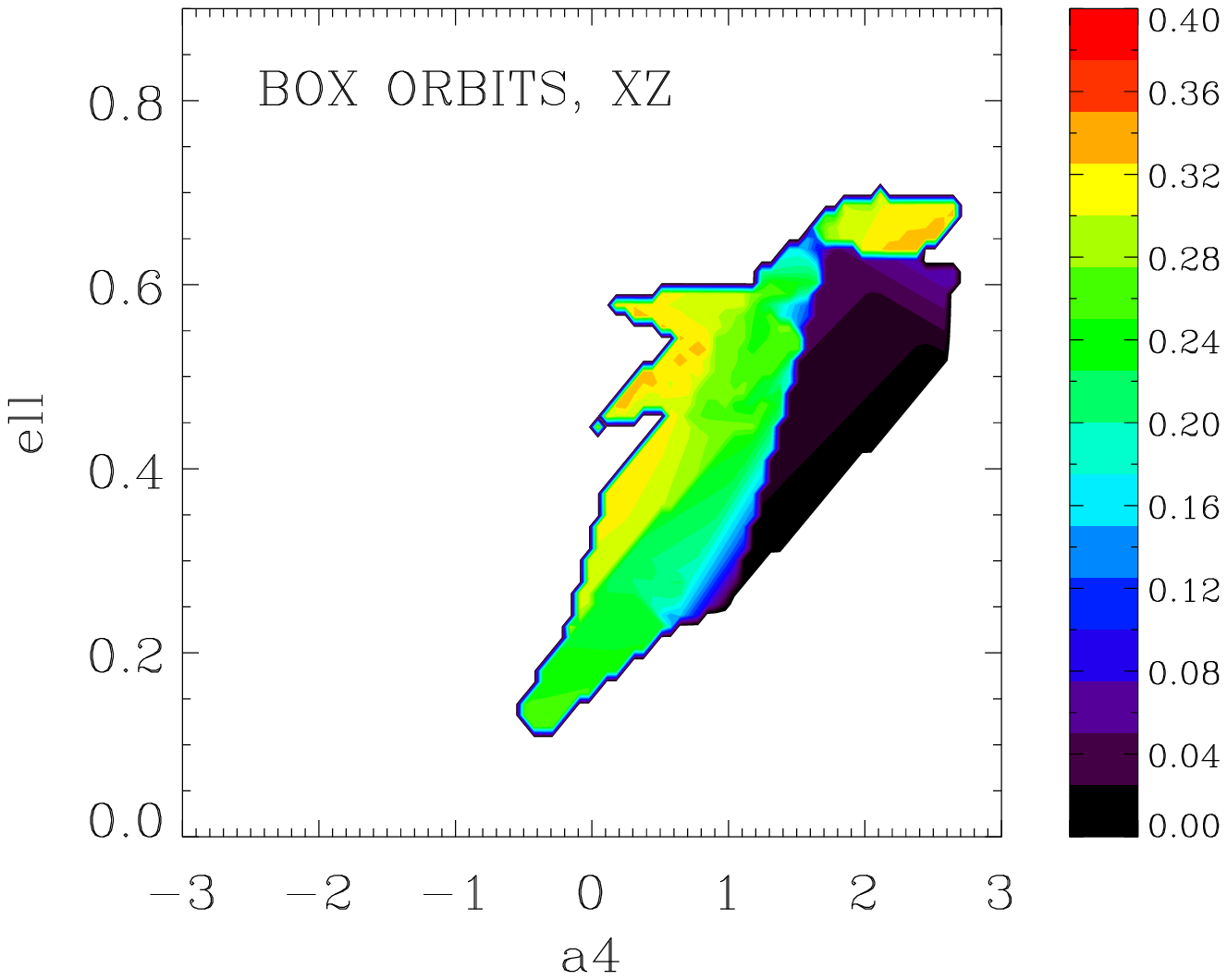}
\includegraphics[angle=0, scale=0.38]{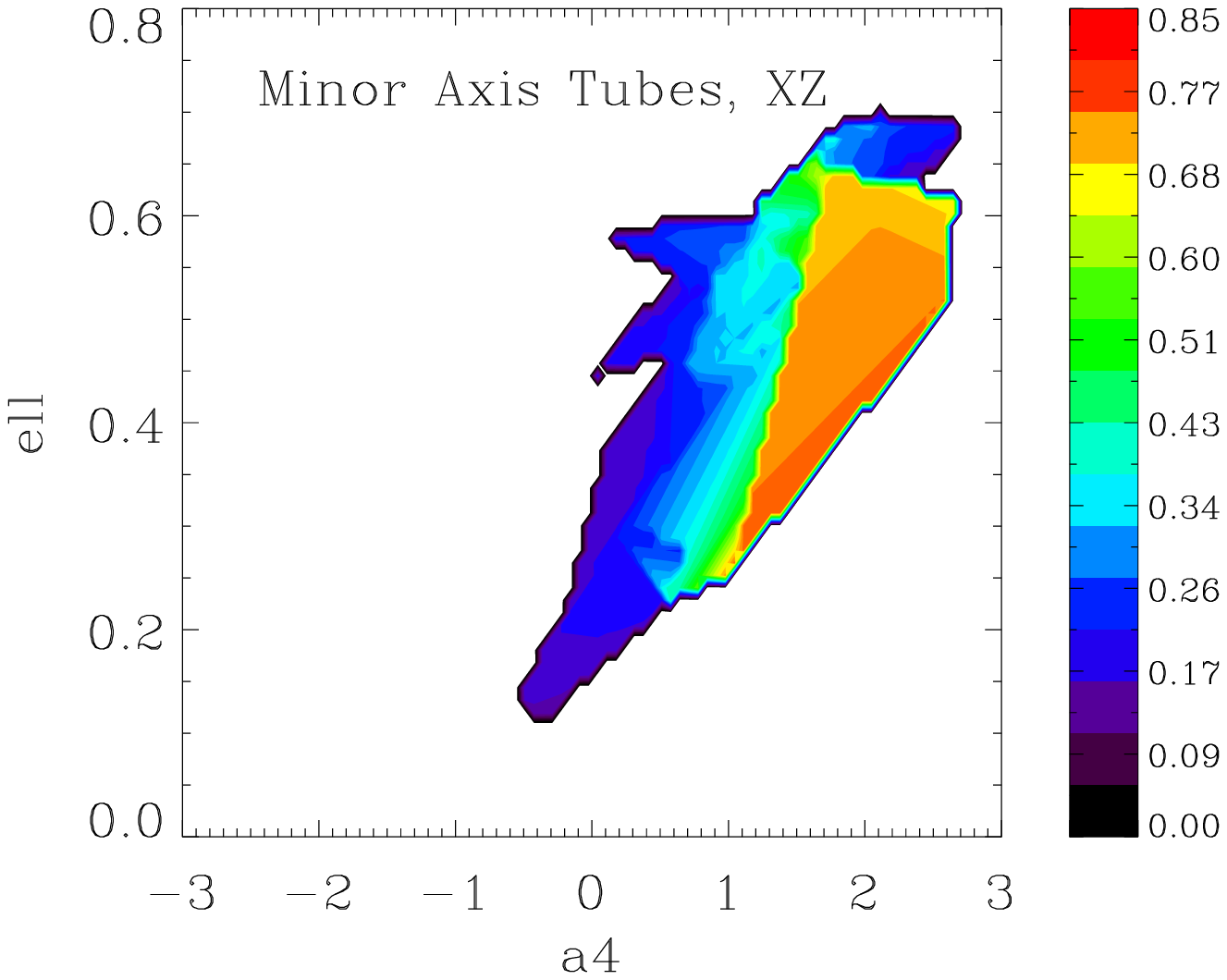}
\includegraphics[angle=0, scale=0.38]{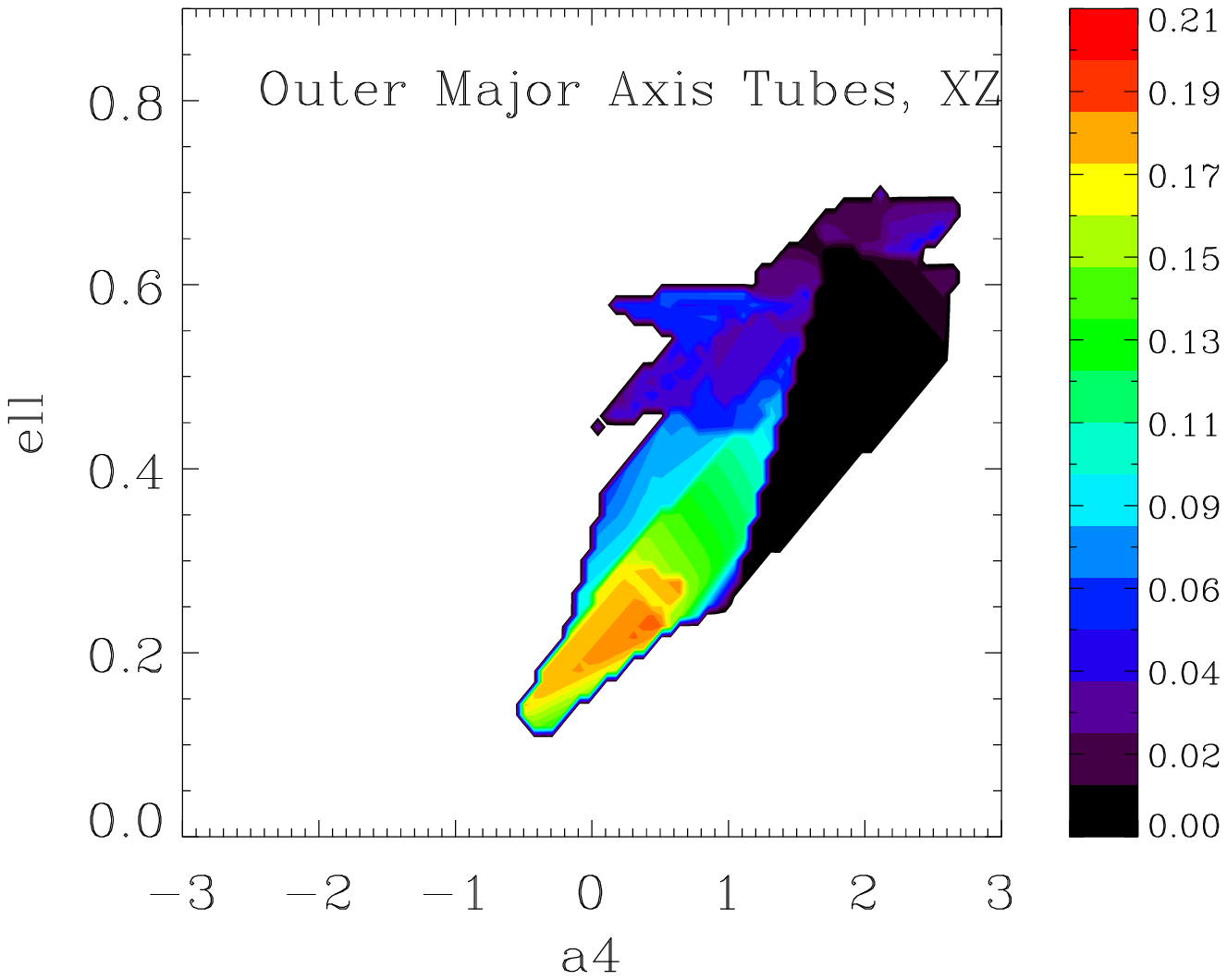}}
\vspace{0.5cm}
\centerline{\includegraphics[angle=0, scale=0.38]{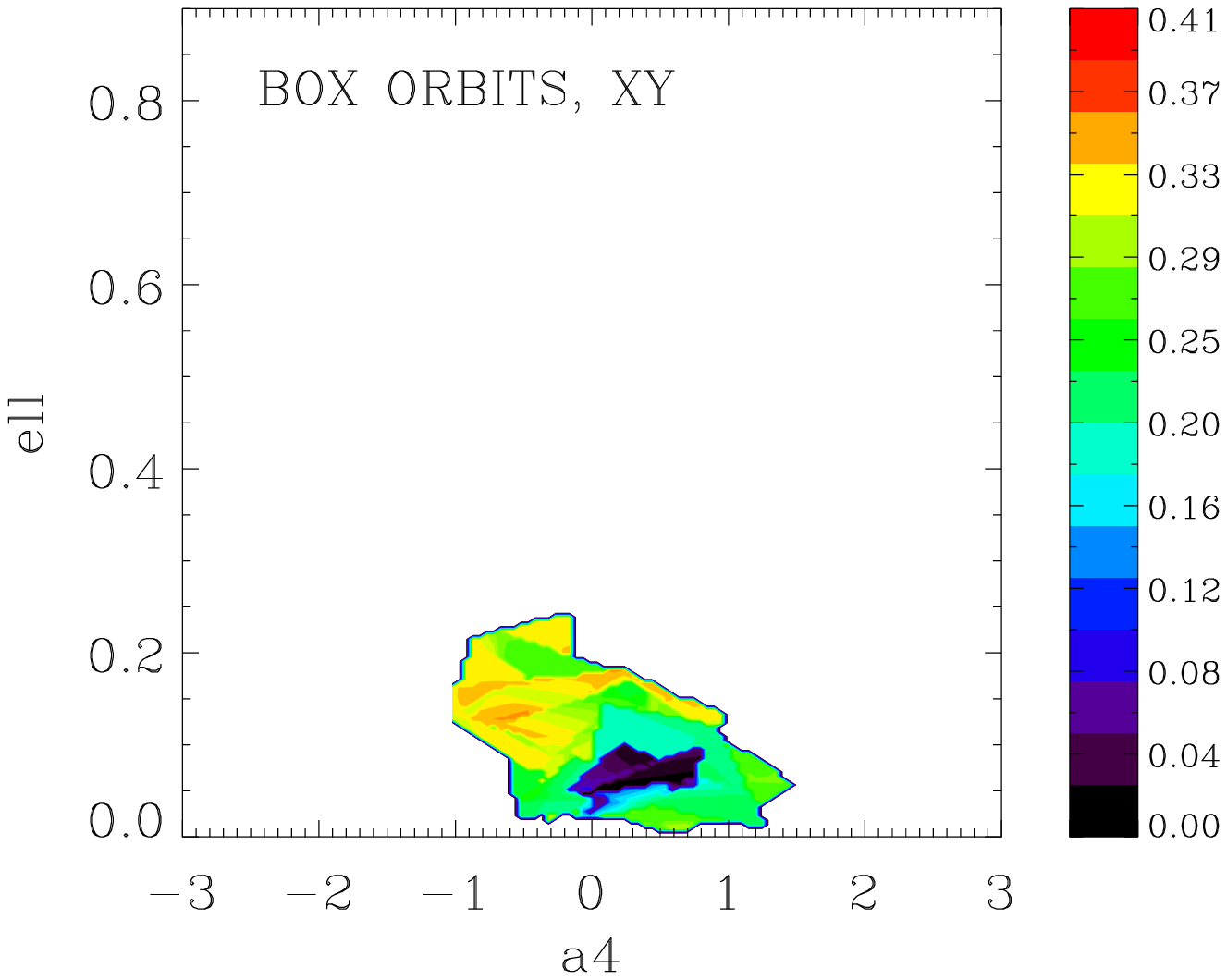}
\includegraphics[angle=0, scale=0.38]{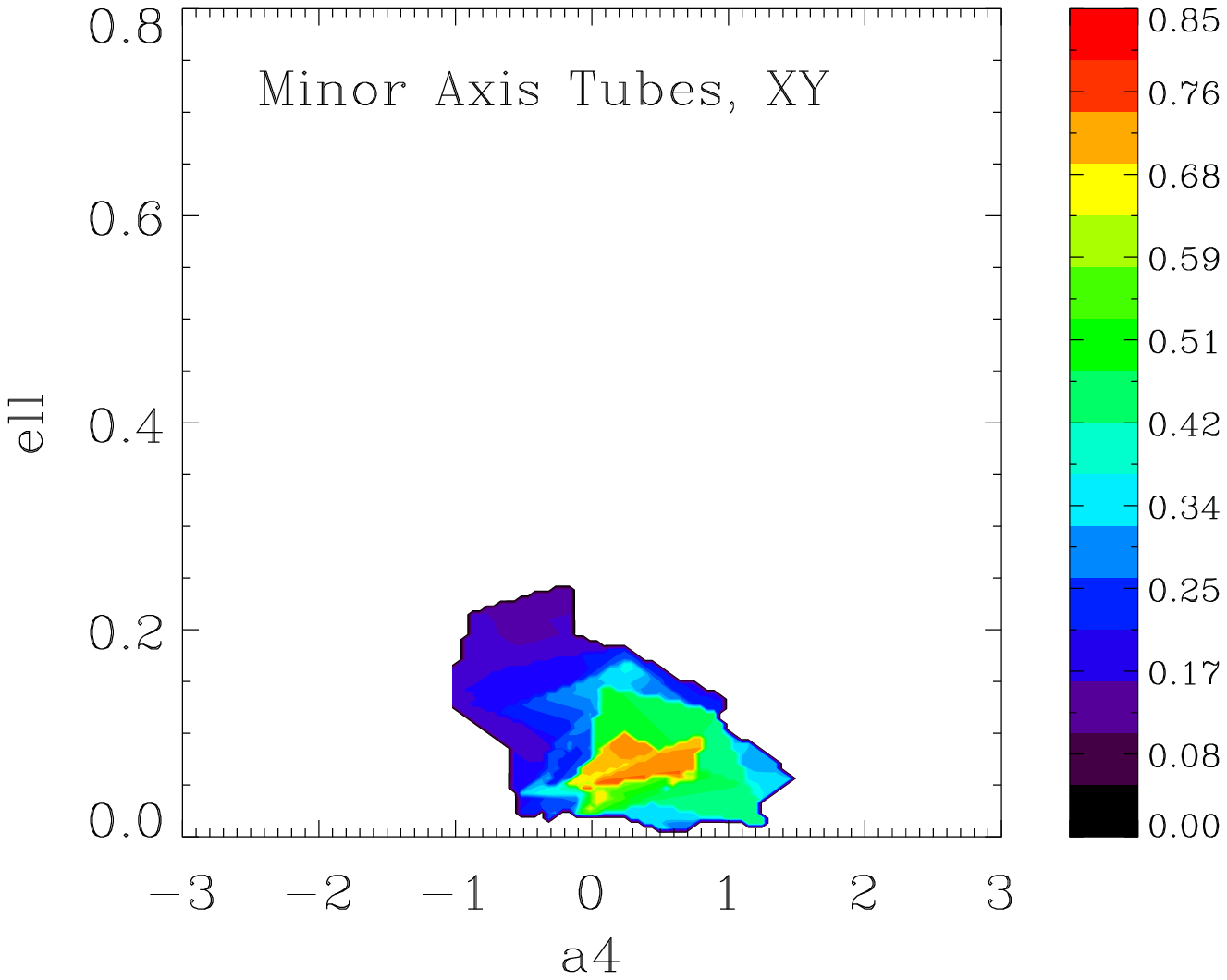}
\includegraphics[angle=0, scale=0.38]{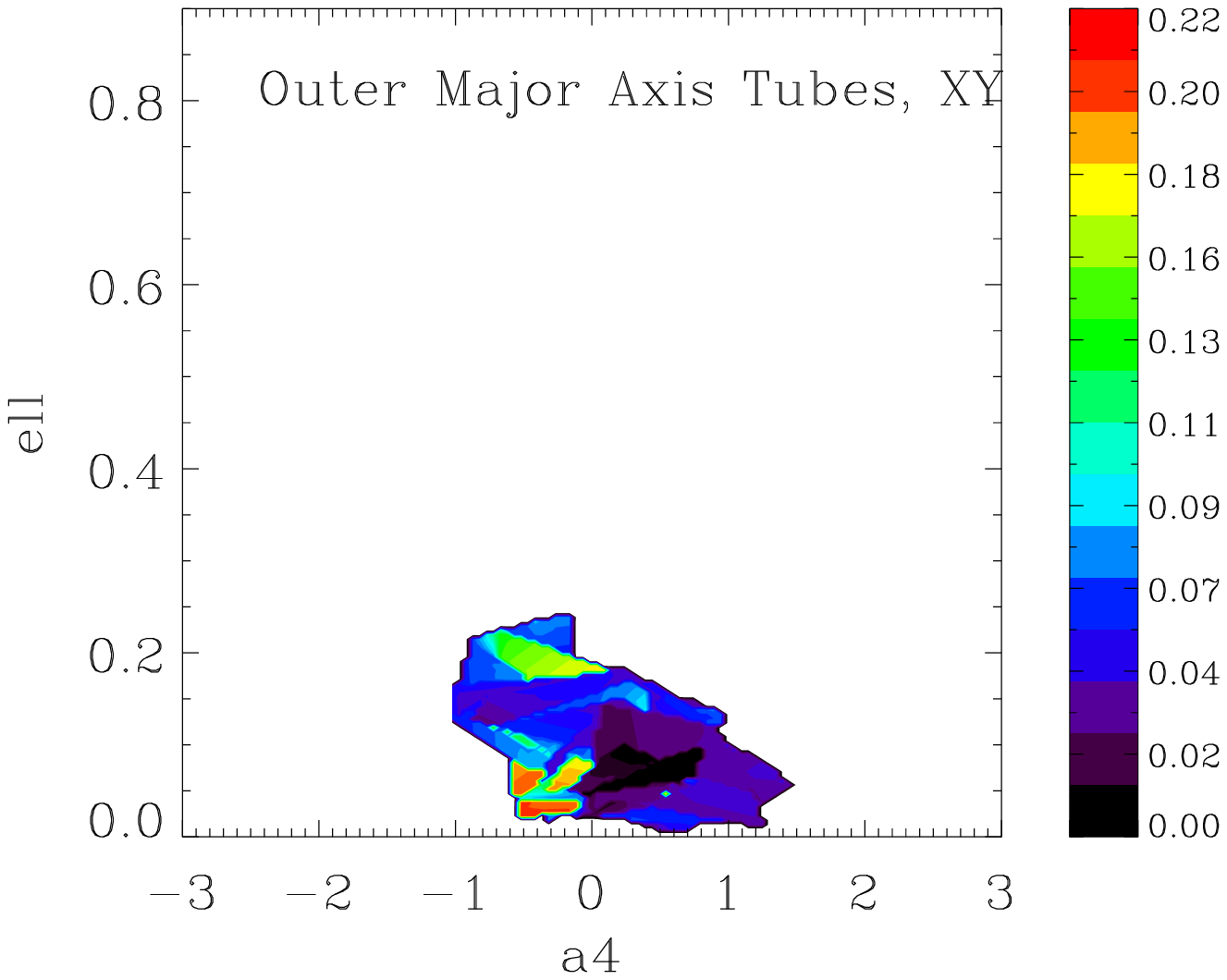}}
\caption{Relation between the photometric properties of the merger remnants,
their intrinsic structure and the viewing angle. Color-coded orbit abundances are shown in
the $a4$-$\epsilon$ plane. In each row we show from left to right the relation the abundance 
of box orbits, minor axis tubes and outer major axis tubes. {\bf Top Row}: Projection along the major axis. 
For this projection the isophotal shape parameter shows the best correlation with the intrinsic structure.
{\bf Middle Row}: Projection along the intermediate axis. This is the most disky projection. The
minor axis tube component can mask the influence of the box orbits effectively.  
{\bf Bottom Row}: Projection along the short axis. All remnants have low ellipticities.}
\label{fig:view}
\end{figure*}

\citet{FRA88} suggested that whether the isophotal shape appears disky or boxy depends
not only on the intrinsic structure of the galaxy, but also on the viewing angle.
This was confirmed in simulations of collisionless disk-disk mergers by \citet{Heyl_94}. As both, the 
intrinsic structure and the viewing angle  are known for the merger remnants it is straight forward
to combine this knowledge. The result is shown in Fig. \ref{fig:view}. To study the influence of the viewing
angle we examine only the projections along the principal axes of the merger remnants, which in almost
all cases result in the most extreme values for $a4$ and $\epsilon$ of all possible projections. 
The projection along the major axis is the most sensitive to the orbital content of the merger remnants. In 
the top row of Fig. \ref{fig:view}, we see that the remnants with negative $a4$ indeed have a dominant
box orbit population, while the most disky ones are dominated by minor axis tubes. Interestingly 
the remnants which have low ellipticities have the highest fraction of outer major axis tubes. The projection
along the intermediate axis (second row) is the most complicated one. Almost none of the remnants are identified
as boxy. Although most of the box orbits dominated remnants now lie close to the $a4=0$ line, some have very disky
projections. Finally the projection along the short axis again is more sensible to the orbital content. The 
box orbit dominated remnants are boxy and the disky remnants are minor axis tube dominated.

Intrinsically the minor axis tubes lead to the most disky shape for short axis projections, while for the
projection along the major axis their shape appears boxy or peanut-like. It seems that a small amount of
minor axis tubes can counter the influence of the box orbits, but only for projections along or close to the 
intermediate axis. This is still enough to generate a large spread in the mean isophotal shape 
as a function of box-to-minor axis tube fraction (Fig. \ref{fig:a4}) for the 3:1 and 4:1 remnants. 

\section{Orbits and Kinematic Properties} 
\label{sec:kin}
The combination of all stars moving on different orbits defines the kinematic properties 
of the remnants. Different orbit classes, however, have different kinematical properties. Minor axis tubes
are responsible for major axis rotation, while major axis tubes are minor axis rotators.
Boxes and boxlets should have a vanishing mean angular momentum. This should result in
correlations between the general kinematical properties of the remnants and their orbital
content. To disentangle these relationships we have extracted the different orbit classes and analyzed their
properties in isolation. This will help us to understand, global projected kinematic properties as well as
more complex phenomena like the line-of-sight velocity distribution (LOSVD).

\subsection{Rotational Support versus Pressure Support}
\begin{figure}
\vspace{1cm}
\centerline{\includegraphics[angle=0,scale=0.38]{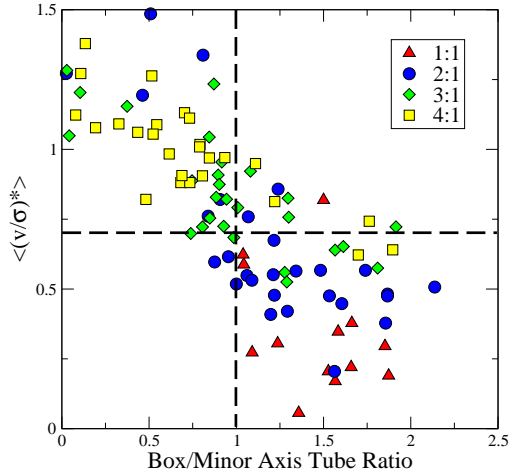}}
\caption{Relation between the anisotropy parameter $(v/\sigma)^*$ and the box to minor axis tube ratio.
The anisotropy parameter is averaged over 50 projections. The horizontal dashed line indicates
$(v/\sigma)^*=0.7$ above which the system is expected to be flattened by rotation.}
\label{fig:aniso}
\end{figure}

The anisotropy parameter $(v_{maj}/\sigma_0)^*$ is defined as the ratio of the observed value
of the rotation along the major axis and the central velocity dispersion, $v_{maj}/\sigma_0$, and 
the theoretical value for an isotropic oblate rotator $(v/\sigma)_{theo}=[\epsilon_{obs}/(1-\epsilon_{obs})]^{1/2}$ 
with the observed ellipticity $\epsilon_{obs}$ \citep{B78}. This parameter has been used by observers to 
determine whether a given galaxy is flattened by rotation $[(v_{maj}/\sigma_0)^*\ge 0.7]$ or by velocity anisotropy
$[(v_{maj}/\sigma_0)^* < 0.7]$ (\citealp{D83}, \citealp{Bender_88}, \citealp{NIETO88}, \citealp{SC95}). 
An intrinsic property of the two most abundant orbit classes is that the minor axis tubes are
dynamically colder and have higher rotational velocities than the box orbits which generate no net rotation and 
which lead to high velocity dispersions. The relative abundance of these orbit classes is connected 
with the value of the anisotropy parameter as can be seen in Fig. \ref{fig:aniso}. 
Interestingly all remnants with $[(v_{maj}/\sigma_0)^*> 0.7]$ are dominated by tube orbits while the majority 
of anisotropic remnants are box orbit dominated. The division line between isotropic and anisotropic remnants 
seems to coincide with a box to tube ratio of unity.

\begin{figure}
\vspace{1cm}
\centerline{\includegraphics[angle=0,scale=0.38]{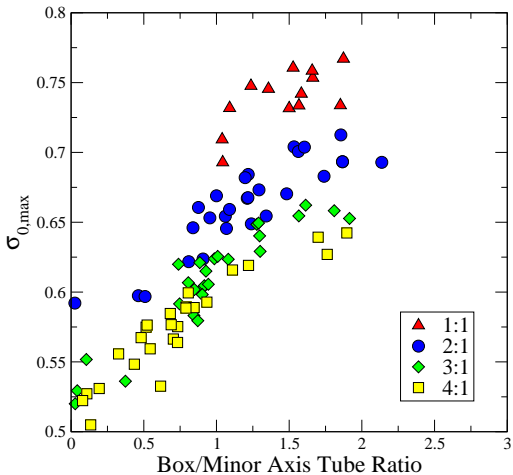}}
\caption{Relation between the maximum central velocity dispersion and box to minor 
axis tube ratio. The projection under which the maximum of the central velocity dispersion is
found is exactly on or very close to the intermediate axis}
\label{fig:sigvrot}
\end{figure}

It is also instructive to examine the relation between velocity dispersion and major axis rotation
directly. The Faber-Jackson relation \citep{FJ} tells us that the central velocity dispersion  
in early type galaxies is tightly connected with the total luminosity (or mass) of the galaxy.
In our sample the merger remnants have four different masses. Equal mass mergers have a total mass
of twice the more massive progenitor, 4:1 mergers have a total mass of 1.25 the more massive progenitor
or 62.5\% the mass of 1:1 mergers. In Fig. \ref{fig:sigvrot} we plot the maximum projected central 
velocity dispersion of each remnant versus the box to minor axis tube ratio. 
The central velocity dispersion $\sigma_0$ of every remnant was determined as the average projected 
velocity dispersion of the luminous particles inside a projected galactocentric distance of 0.2 $r_\mathbf{eff}$.
As expected more massive remnants have higher central velocity dispersions. What 
is more remarkable is that, for a given mass, there exists a correlation with the
box to minor axis tube ratio. The conclusion must be that the central velocity 
dispersion is also an indicator of initial conditions (in this case merging symmetry), which 
populate the different orbit classes accordingly. Unfortunately, if projection effects are taken into account,
the correlation becomes less clear. It is hard to think of a way to use this property in observations
of real ellipticals.\\

\subsection{Minor Axis Rotation}
\begin{figure}
\vspace{1cm}
\centerline{\includegraphics[angle=0,scale=0.38]{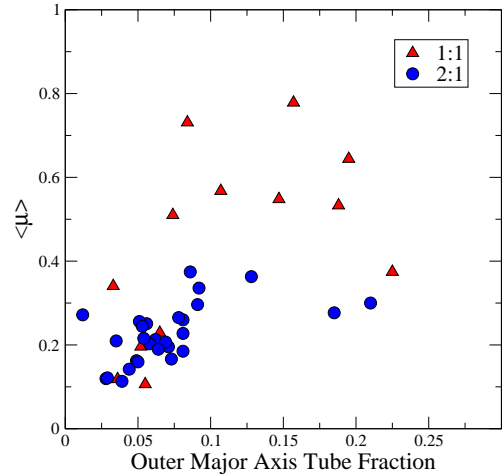}}
\hspace{0.75cm}
\caption{Relation between $\mu=v_{min}/({v_{maj}}^2+{v_{min}}^2)^{1/2}$ 
averaged over 50 projections and the outer major axis tube fraction. The $\mu$-parameter
is used in observational analysis to determine the amount of minor axis rotation
in a galaxy.}  
\label{fig:mu}
\end{figure}

The amount of minor-axis rotation can be parameterized as 
$\mu=v_{min}/({v_{maj}}^2+{v_{min}}^2)^{1/2}$ \citep{B85}. It is an important 
indicator in combination with the isophotal twist for triaxiality in
elliptical galaxies (\citealp{WAG88}, \citealp{FRA91}). Only the 1:1 and 2:1 merger 
remnants have significant minor axis rotation (see also Fig. 13 of \citealp{NB03}). In Fig.
\ref{fig:mu} we plot $\mu$ versus the outer major axis tube fraction.
$\mu$ is measured at $0.5 r_\mathbf{eff}$ and averaged over 50 projections. 
In general, mergers with a higher outer major axis 
tube fraction have a higher value of $\mu$. Although this result is encouraging, there 
are some outliers with low minor axis rotation, but high outer major axis tube 
fraction. This could be explained by counter-rotating populations, which lower 
the minor axis rotation.

\subsection{Line-of-Sight Velocity Distributions}
\subsubsection{Velocity Profiles}
\begin{figure*}
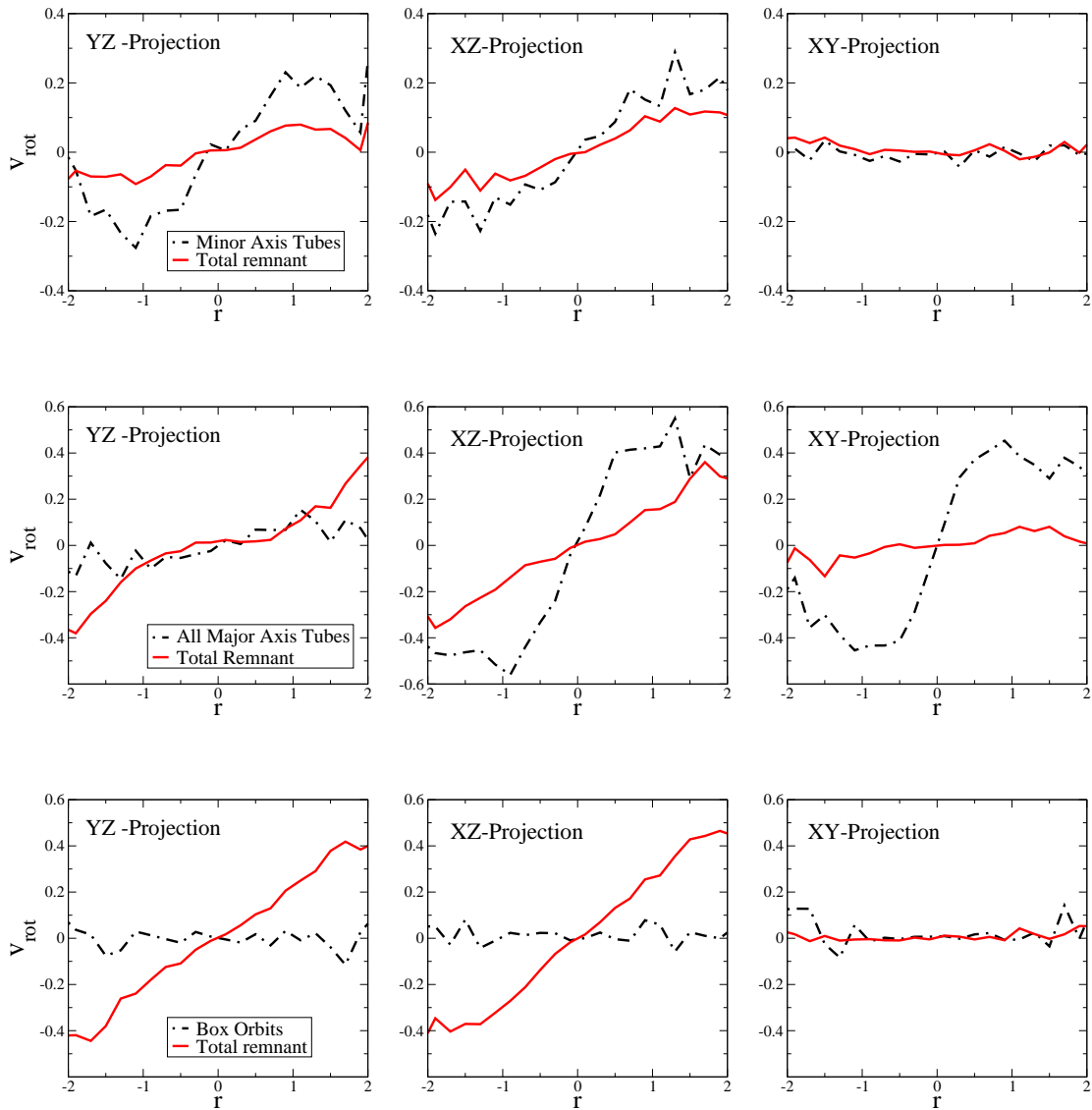

\vspace{1cm}
\centerline{\includegraphics[angle=0, scale=0.5]{vrot_ztube_11MCS_09.eps}}
\vspace{1cm}
\centerline{\includegraphics[angle=0, scale=0.5]{vrot_totxt_11MCS_07.eps}}
\vspace{1cm}
\centerline{\includegraphics[angle=0, scale=0.5]{vrot_box_31MCS_13.eps}}
\caption{Comparison of the rotational properties of single orbit classes (dot-dashed lines) with
the rotation of the whole remnant (solid lines). Three characteristic remnants, two 1:1 and one 3:1, are chosen.
In each case an orbit class is extracted and projected along the major, intermediate and 
short axis (YZ, XZ, XY-plane). The observational slit is positioned along the apparent 
major axis.{\bf Top Row}: Minor axis tubes in a 1:1 remnant. 
They show rotation in the YZ and XZ projection and no rotation in the XY projection.
{\bf Middle Row}: Major axis tubes (both types) in a prolate 1:1 remnant. Strong rotation is present
in the XZ and the XY projection and almost non in the YZ projection. {\bf Bottom Row}: 
Box orbits in a 3:1 remnant. They show no rotation in any projection as expected, however the whole remnant
does rotate strongly.}
\label{fig:vprof}
\end{figure*}

A simple test can show whether the particles which we have assigned to a certain orbit class show
the typical kinematical behaviour that we would expect. 
In Fig. \ref{fig:vprof} the mean line-of-sight velocity (LOSV), along the three principal
axes is shown. The measurement of the total merger remnant is compared with the signature which one
would get by just analyzing the subset of particles, classified as minor axis tubes, major axis tubes 
and box orbits. The minor axis tubes show a higher rotation velocity for the two projections along the 
intermediate and the long axis (XZ and YZ) than the total remnant and show no rotation for the face on 
(XY) projection. Similarly in the same figure, second row, the major axis tubes show significant rotation 
for the XZ and the XY projection and almost no rotation for the YZ projection where we observe this orbit 
class face on. Note that the XZ projection has contributions from both tube types, the major and the minor 
axis tubes. This is the reason, why the major axis tubes dominate more for the XY projections, 
where there is no contribution from the minor axis tubes. Finally the box orbit population in 
a 3:1 remnant does not show any rotation, as we would expect, 
although the total remnant has a high amount of rotation. 

\subsubsection{Global Correlations}
\begin{figure*}
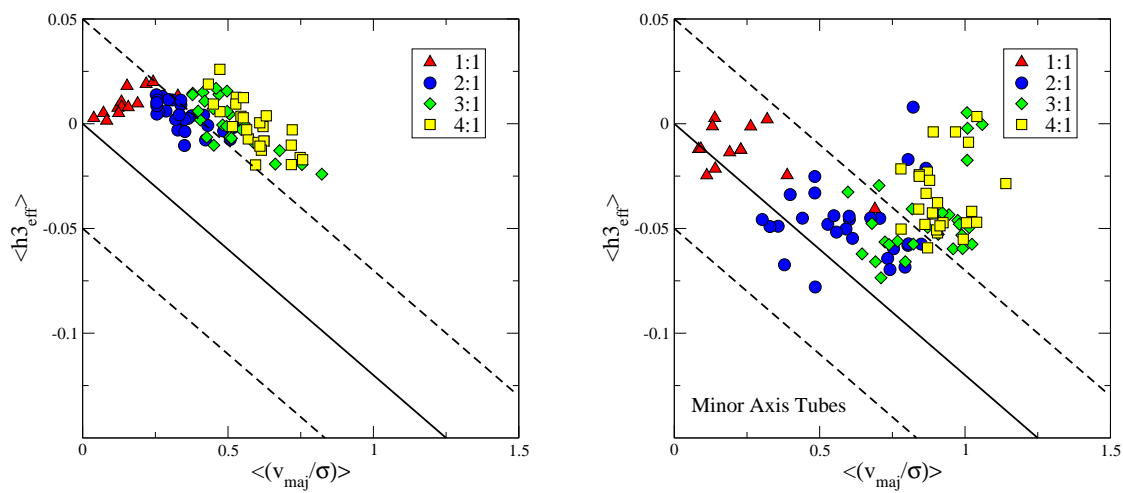

\vspace{1cm}
\centerline{\includegraphics[angle=0,scale=0.38]{correl_h3eff_vsigma.eps}
\hspace{0.75cm}
\includegraphics[angle=0,scale=0.38]{correl_h3eff_vsigma_ztube2.eps}}
\caption{The ratios of rotation along the apparent major axis and central 
velocity dispersion as function of the effective $h3$ parameter are shown for simulated
merger remnants and compared with observed
galaxies as reported by \citet{BSG94}. The observed correlations are indicated by solid lines, the
observational spread by dashed lines.
{\bf Left:} Measurements for the total remnant, each remnant is averaged over 50 projections. The values
are far too positive {\bf Right:} Only minor axis tubes, also averaged over 50 projections. The minor axis tube
components of 1:1 and 2:1 merger remnants agree well, at least kinematically, with observations. 3:1 and
4:1 mergers also have more negative values, but not enough to explain the observational
data.}
\label{fig:bender}
\end{figure*}

\begin{figure*}
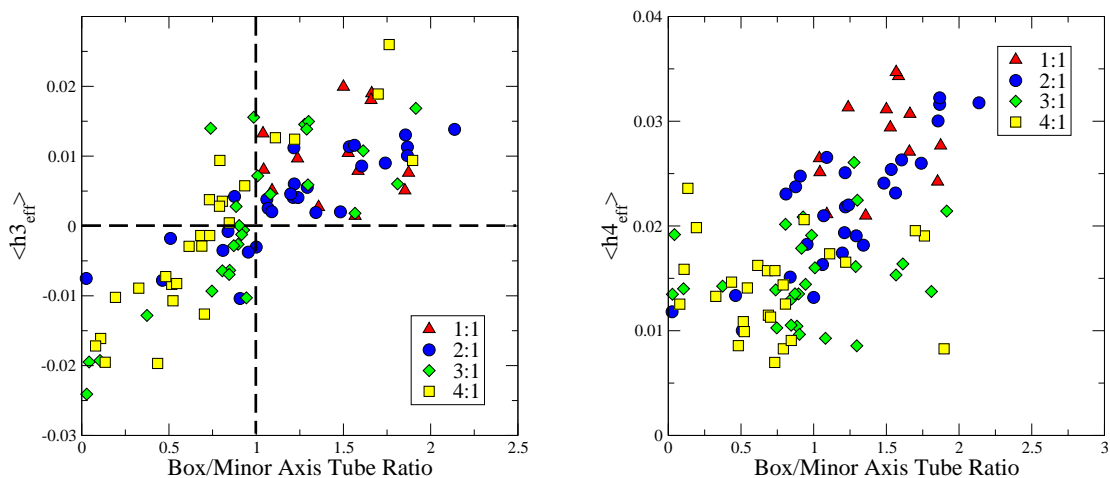

\vspace{1cm}
\centerline{\includegraphics[angle=0,scale=0.38]{correl_h3eff_btoz.eps}
\hspace{0.75cm}
\includegraphics[angle=0,scale=0.38]{correl_h4eff_btoz.eps}}
\caption{{\bf Left:} Relation between the box to minor axis tube ratio and $h3_\mathbf{eff}$. The
horizontal line divides positive and negative values for $h3$ and the vertical line
minor axis tube from box orbit dominated remnants. Box orbit dominated remnants tend
to have positive $h3_\mathbf{eff}$.
{\bf Right:} Relation between the box to minor axis tube ratio and $h4_\mathbf{eff}$ }
\label{fig:h3h4}
\end{figure*}

It is known that the LOSVDs of early-type galaxies can deviate significantly
from pure Gaussian profiles. The two most important parameters are the third order coefficient of the 
Gauss-Hermite expansion, $h3$, which measures the asymmetric deviations and the fourth order coefficient
$h4$  which measures the symmetric deviations from a Gaussian profile. We follow the 
definition of \citet{BSG94} in calculating the effective parameters, termed $h3_\mathbf{eff}$ and 
 $h4_\mathbf{eff}$ by averaging from the center of the remnant to 0.75 $r_\mathbf{eff}$. They found 
correlations between $h3_\mathbf{eff}$ and  $(v/\sigma)$ as indicated in Fig.\ref{fig:bender}. The 
merger remnants cannot reproduce this correlation. It is apparent
that the values scatter around zero for $h_3\mathbf{eff}$ and are not as negative as expected from
observations for a given $v/\sigma$. The picture
changes if the same analysis is done taking only particles on minor axis tubes. Much more
negative $h3_\mathbf{eff}$ can be reached. The measured values for the 1:1 and the 2:1 remnants 
even fall on the observed relation, the 3:1 and 4:1, however, do not. While
the merging process is sufficiently violent to introduce asymmetries in the LOSVD 
in the 1:1 and 2:1 remnants, it is not so in the 3:1 and 4:1 remnants.
The problem remains that the $h3_\mathbf{eff}$ is too positive for the {\it total} merger remnant.
One must assume that the other orbit classes have an impact, too. This is hinted at in 
Fig.\ref{fig:h3h4}, left panel. We see that merger remnants with positive h3 values tend to have a 
dominant box orbit population and we do not find a single box-dominated remnant with a negative 
$h3_\mathbf{eff}$. Even some tube-dominated subsamples, like the 4:1 mergers, show a tight correlation 
of $h3_\mathbf{eff}$ with the box to minor axis tube ratio. The situation for $h4_\mathbf{eff}$ 
is not as clear. Although some correlation can be seen for very box rich remnants, there is no evidence
for this in 3:1 and 4:1 remnants. $h4_\mathbf{eff}$ is positive in all remnants.

\subsubsection{Local Correlations}
\begin{figure}
\vspace{1cm}
\centerline{\includegraphics[angle=0,scale=0.38]{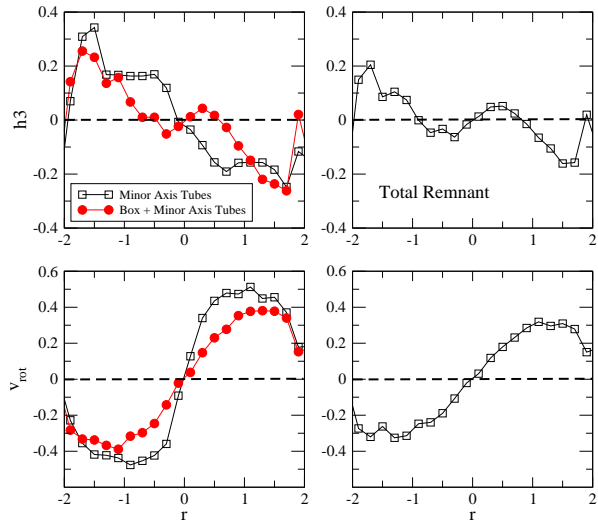}}
\caption{{\bf Left:} $h3$ (top) and $v_{rot}$ (bottom) for a 3:1 remnant taken along the
major axis for minor axis tubes and minor axis tubes and box orbits superposed.
While for the minor axis tubes the signs of  $h3$ and $v_{rot}$ are anti-correlated,
a correlated inner ridge forms in $h3$ when the box orbits are added.
{\bf Right:} Like before, but for the total remnant}
\label{fig:local}
\end{figure}

In addition to observed correlations for the effective $h3$ there also exist local correlations 
within a single galaxy. 
As expected from a rotating axisymmetric system the rotation velocity is anti-correlated with $h3$ 
and has opposite sign (\citealp{BSG94}; \citealp{HALL01}; \citealp{PINK03}). This anti-correlation 
is violated in most collisionless merger remnants (\citealp{BB_00}, \citealp{NB01}) and poses a serious
problem for the merger hypothesis. Decomposing the merger remnant into orbital components
is an ideal tool to shed light on this question. Fig. \ref{fig:local} shows a 
3:1 merger remnant, observed along the true major axis. The $h3$ and $v_{rot}$ for the total remnant are 
anti-correlated in the outer parts, but correlated in the center (Fig. \ref{fig:local}, right side). 
Exactly the same procedure is done with the particles classified as minor axis tubes. This time 
an anti-correlation is seen over the whole radial range. It is only when we superpose the box 
orbits that the kinematics starts to resemble closely the kinematics of the whole remnant. We find 
that the minor axis tube components of {\it all} merger remnants display the 
observationally predicted $h3$-$v_{rot}$
anti-correlation. One could conclude that there can not exist a sizeable box orbit population in the
center of real elliptical galaxies. But probably the situation is more complicated in
real galaxies, e.g. if a central disk component is present \citep{NB01}.
Also in our 1:1 remnants the sometimes strong major axis tube component can complicate the
effect of superpositions considerably. 

\section{Summary and Conclusions}
\label{sec:end}
We have presented a detailed analysis of the orbital content of a statistical sample 
of collisionless merger remnants with mass ratios of 1:1, 2:1, 3:1, and 4:1 
using the spectral method of CA98. The orbital content has been tested to remain 
constant over at least a few dynamical times at the half mass radius. For the 1:1 sample 
we have repeated the analysis with remnants of bulgeless progenitor disks and remnants 
of mergers with a larger pericenter distance. We have investigated in detail the relation between
the orbital content of the merger remnants without figure rotation 
and their projected kinematic and photometric properties. For remnants with bulges
 minor axis tubes and box orbits are the dominant orbit class, independent of the mass ratio. 
Only for remnants of bulgeless disks, boxlets and box orbits dominate the systems. 
On average the fraction of minor axis tubes doubles from a mass ratio of 1:1 to 4:1 
whereas the fraction of box orbits decreases weakly. The minor 
axis tube orbits predominantly originate from the progenitor disks. 
In contrast equal numbers of box orbits stem from the disk and the bulge
of the progenitor galaxies. The triaxiality of the remnants correlates 
with the orbital content. The minor axis tube fraction strongly decreases 
from oblate to prolate systems while the remnants with the highest box orbit fraction 
are the most triaxial ones. This is in agreement with analytical predictions for self consistent 
triaxial systems \citep{STAT87}. 

Interestingly, the division line defined by observers between pressure supported 
and rotation supported systems of $(v/\sigma)^* =0.7$ coincides with a box to 
minor axis tube ratio of unity. Pressure supported systems have more box than minor
axis tube orbits, minor axis tubes prevail in rotation supported systems. Strong minor axis 
rotation can, as expected, only be found in remnants with a significant number of major 
axis tube orbits. The projected central velocity dispersion of the remnants 
is not only connected to their mass but also to their balance of box to 
minor axis tube orbits. 
 
The observed anti-correlation between the asymmetry of the LOSVD measured by $h_3$  
and the ratio of $v/\sigma$  found in early-type galaxies by \citet{BSG94} can not be 
reproduced by collisionless merger remnants. We do find some remnants with negative 
$h3_{\mathbf{eff}}$  but with an amplitude that is a factor of five too small 
compared to observations. All remnants with more box than minor axis 
tubes show a positive $h3_{\mathbf{eff}}$. The observed correlation can be better reproduced
if we consider only particles on minor axis tubes. Therefore we conclude that 
any additional physical process during the merging of the galaxies that would reduce the number 
of box orbits and increase the number of tube orbits  would naturally lead to the 
observed correlation. The most likely process is the destruction of box orbits by a 
large mass concentration at the center of the galaxies \citep{LEES92}. Those mass concentrations 
can arise from infalling gas during a merger of gas rich disks \citep{BH_96}. Alternatively 
a very concentrated bulge could have a similar effect. 

Merger remnants which are dominated by minor axis tube orbits have predominantly disky 
projections. However, they can appear boxy for certain projections. It is only possible 
to produce a pure boxy remnant when the majority of particles move on box orbits. 
This might provide a fundamental limitation for the applicability of axisymmetric 
Schwarzschild modeling of the most massive, probably triaxial, elliptical galaxies 
\citep{Cretton_00}. These models can only reproduce boxy isophotal shapes with tube orbits; 
our models predict that those galaxies will on average have disky projections.  

\section*{Acknowledgments}
We are grateful to Daniel Carpintero and Luis Aguilar for making available their 
classification code and to Shunsuke Hozumi for making available his self consistent field code. 
We thank Ralf Bender, James Binney, Hans-Walter Rix for helpful discussions and comments. 
Roland Jesseit acknowledges financial support by the SFB 375 Astro-Teilchenphysik of the DFG.

\label{lastpage}

\end{document}